 \documentclass[format=acmsmall, review=false, screen=true, nonacm]{acmart}
 \setcopyright{none}

\usepackage{enumerate}
\usepackage{booktabs}
\usepackage{soul}
\usepackage{eurosym}
\usepackage{todo}

\usepackage[utf8]{inputenc}

\usepackage{booktabs} 

\begin{document}
\title{On the mapping of Points of Interest through StreetView Imagery and paid crowdsourcing}

\author{Eddy Maddalena}
\orcid{1234-5678-9012}
\affiliation{%
  \institution{University of Southampton}
  \streetaddress{P.O. Box 1212}
  \city{Southampton}
  \country{United Kingdom}
  \postcode{SO17 1BJ}
}
\email{e.maddalena@southampton.ac.uk}

\author{Luis-Daniel Ibáñez}
\orcid{1234-5678-9012}
\affiliation{%
  \institution{University of Southampton}
  \streetaddress{P.O. Box 1212}
  \city{Southampton}
  \country{United Kingdom}
  \postcode{SO17 1BJ}
}
\email{l.d.ibanez@southampton.ac.uk}

\author{Elena Simperl}
\orcid{1234-5678-9012}
\affiliation{%
  \institution{University of Southampton}
  \streetaddress{P.O. Box 1212}
  \city{Southampton}
  \country{United Kingdom}
  \postcode{SO17 1BJ}
}
\email{e.simperl@southampton.ac.uk}

\renewcommand\shortauthors{Maddalena, E. et al}

\begin{abstract}
The use of volunteers has emerged as low-cost alternative to generate accurate geographical information, an approach known as Volunteered Geographic Information (VGI). However, VGI is limited by the number and availability of volunteers in the area to be mapped, hindering scalability for large areas and making difficult to map within a time-frame. Fortunately, the availability of street-view imagery enables the virtual exploration of urban environments, making possible the  recruitment of  contributors not necessarily located in the area to be mapped. In this paper, we describe the design, implementation, and evaluation of the Virtual City Explorer (VCE), a system to collect the coordinates of Points of Interest within a bounded area on top of a street view service with the use of paid crowdworkers. Our evaluation suggests that paid crowdworkers are effective for finding PoIs, and cover almost all the area. With respect to completeness, our approach does not find all PoIs found by experts or VGI communities, but is able to find PoIs that were not found by them, suggesting complementarity. We also studied the impact of making PoIs already discovered by a certain number of workers \emph{taboo} for incoming workers, finding that it encourages more exploration from workers , increase the number of detected PoIs , and reduce costs.
\end{abstract}

%
%

%
%


\maketitle

\section{Introduction}
\label{sec:intro}
As national mapping agencies came under funding pressures, Volunteered Geographic Information (VGI)~\cite{sui_crowdsourcing_2012} arose as a low-cost alternative to generate maps. VGI takes advantage of the lowered technological barrier for common citizens, \emph{i.e.}, the availability of low-cost GPS sensors embedded in mobile phones, and open source geographical information systems, to enable volunteers to contribute to the collection and generation of geographic information in the areas they live and frequent~\cite{goodchild_sharing_2007}. A number of studies (e.g. \cite{haklay_openstreetmap:_2008,fan_quality_2014,dorn_quality_2015}) have assessed the quality of VGI with respect to mapping experts, concluding that in general it achieves high completeness and correctness (greater than $80\%$), with a fair positional accuracy (4-8 meters of difference). The same lowered technological barrier has been exploited for audit and monitoring of urban infrastructure, urban health and accessibility, empowering inhabitants of urban areas to contribute while improving their sense of community and their relationship with public authorities \cite{ruiz-correa_sensecityvity:_2017,jiang_citizen_2016}

However, a recognized limitation of VGI is its dependence on the number of volunteers that can physically move to the area to be mapped, and their level of engagement~\cite{quattrone_work_2017}. VGI contributors have been observed to be more likely to contribute in areas geographically close to them~\cite{hardy_volunteered_2012}, while the positional accuracy of contributions has been positively correlated with the number or contributors involved~\cite{haklay_how_2010}. Furthermore, the number of contributions is not linear, and their location is generally unpredictable~\cite{mooney_analysis_2014}, making difficult to assess if specific sectors within the area of interest have been covered, and to generate maps within a specific time-frame. Fortunately,
recent advances in street-view imagery have made possible the collection and publication of detailed \emph{virtual} snapshots of real-world areas, like the popular Google Street View service~\cite{anguelov_google_2010}. These \emph{virtual spaces} can be navigated with personal devices in almost the same way as real spaces, opening the door to contributions from people not necessarily located in the area to be mapped, increasing the scalability of the approach. In particular, paid crowdworkers could be recruited on-demand to produce geographic information, or perform an urban auditing. 

Previous efforts to use street-view imagery for collecting geographical information have focused on urban auditing tasks where workers either walk a pre-determined itinerary (\emph{e.g.}, street segments like in~\cite{saha_pilot_2017}) or are teleported to a previously known point of interest in the virtual space, where they perform the task (\emph{e.g.} audit the state and accessibility of a bus stop~\cite{hara_improving_2015}). However, there are no previous studies on the applicability of street-view imagery for \emph{exploratory} tasks, \emph{i.e.}, where instead of being asked to go to specific points in space, humans are required to locate all the Points of Interest (PoIs) in a given area. Exploratory tasks are more time-consuming than \emph{go-to-point} ones for urban auditors or volunteers on the field, as the whole area of interest needs to be explored to be certain that all PoIs have been located.


In this paper, we describe the design, implementation and evaluation of the Virtual City Explorer (VCE), a system to collect the coordinates of any PoI within a bounded area of interest on top of a street view service. The VCE enables end users (e.g., mapping agencies, municipalities, citizen collective) to configure the areas of interest, target PoIs, and number of crowdworkers to be recruited. With these parameters, the VCE generates a Web interface that enables crowdworkers the virtual exploration of the area of interest in a street-view service, and the reporting of PoIs' coordinates. To consolidate coordinates sent by different workers that may correspond to the same PoI, our system uses the DBSCAN algorithm to cluster them, producing the final map.


We evaluated the VCE in two areas of two different cities: (1) The \emph{Limited Traffic Zone} of Trento, Italy, where we compared against a map generated by experts of the City Council going on the field, and data from OpenStreetMaps; and (2) a custom area of Washington D.C., USA, where we compared against a map generated by a very active VGI community focused on bike racks (Rackspotter), and data from OpenStreetMaps. Our system pursues the following research questions:

\begin{enumerate}[{RQ1.}]
\item \label{item:rq1} Feasibility and completeness: How accurately paid crowd workers working on street-view imagery locate PoIs? Are they able to locate all the PoIs within an area?

\item \label{item:rq2} 
Area coverage: Do crowdworkers cover all the area of interest? 

\item \label{item:rq3} Cost and time: How does our approach compares with approaches requiring physical presence in terms of cost and number of discovered PoIs?

\end{enumerate}

During the experiments, we observed that a number PoIs were independently discovered by a large number of workers, potentially wasting workers' effort, and end user budget. In response, we implemented the following optimisation: When a PoI has been detected by a predetermined number of workers, or intuitively, independently confirmed a number of times, it becomes \emph{taboo}, \emph{i.e.}, it is not possible for other workers to report it anymore, prompting them to further explore the area. To avoid being unfair to workers that start the task when most PoIs have already been detected, we also reward workers who did not report the requested number of PoIs, but spent a certain amount of time in the task, covering a certain  distance. This leads to a further research question:
\begin{enumerate}[{RQ4.}]
\item  \label{item:rq4} Does the \emph{taboo} optimisation improves area coverage and completeness of our approach? 
\end{enumerate}

Finally, to advance towards the understanding of the right number of workers needed to map an area, and where the VCE interface needs to be improved, we ask a number of research questions aimed to investigate the workers' interaction with the VCE, that we group as follows:



\begin{enumerate}[{RQ5.}]

\item \label{item:rq5} 


 How much time do workers spend on the task and how much distance do they cover? Does the taboo strategy affect time spent and distance walked? Which are the most frequent errors that affect the user experience? 

\end{enumerate}

The remainder of the paper is organised as follows, Section \ref{sec:relwork} presents an overview of related work. Section \ref{sec:systemdesign} describes our system and the interface that workers use. Section \ref{sec:experimentdesign} describes our experiments and the results observed without the Taboo optimisation, in connection with RQs 1, 2, and 3. Section \ref{sec:taboo} introduces the Taboo optimisation and compares it with the not optimised version, in connection with RQ4. Section \ref{sec:behaviour} analyses the behaviour of workers when working under the basic and the optimised strategy, in connection with RQ5. Finally, section \ref{sec:conclusions} summaries our findings and outlines areas of future work.

This work was approved by the Ethics and Research Governance (ERGO II) committee of the Faculty of Engineering and Physical Sciences (FEPS), University of Southampton, on 16 April 2018, with submission ID 41038.



\section{Related Work}
\label{sec:relwork}
Volunteered Geographical Information (VGI) is a bottom-up method based on the participation of volunteers to generate maps, as opposed to top-down methods involving experts~\cite{sui_crowdsourcing_2012}. The most known example of VGI is probably the OpenStreetMap project,\footnote{\url{https://www.openstreetmap.org}, accessed on November 23rd 2018.} an open, collaborative initiative to produce a highly detailed world map, editable by anyone~\cite{haklay_openstreetmap:_2008}. \cite{haklay_openstreetmap:_2008,fan_quality_2014,dorn_quality_2015} have observed that maps generated with VGI have high completeness and correctness ($> 80\%$), with a fair position accuracy (4-8 meters of difference) with respect to maps generated by experts from mapping agencies. The main limitation of VGI approaches is that quality is proportional to the number of contributors who have worked on a given spatial unit ~\cite{haklay_how_2010}. \cite{hecht_localness_2010} introduced the notion of \emph{Spatial Content Production models} to categorize the localness of VGI-generated content, highlighting the differences between (i) \emph{you have to be there} and (ii)\emph{flat earth} models. In (i) contributors are required to be physically present in a certain area of interest, while in (ii) the distance between contributor and area of interest is not relevant, \emph{i.e}, a contributor located anywhere on Earth can contribute information about a point located anywhere on Earth. In contrast with the traditional VGI approach that fits (i), our approach enables a PoI mapping model that conforms to (ii). By allowing the recruitment of contributors not located in the area of interest, our approach increases their number per spatial unit, therefore increasing the quality of generated maps.

Crowdsourcing tasks that require workers to move in space to perform them are studied in the field of Spatial Crowdsourcing (SC)~\cite{zhao_spatial_2016}. VGI can be seen as a special case of SC, where contributors are volunteers and the goal of the task is to collect or verify geo-spatial information. As VGI, SC faces challenges regarding availability of workers in areas where they are needed, plus specific challenges like optimal task assignment according to workers' current positions and available budget, and how to preserve their privacy~\cite{kandappu_obfuscation_2018,tong_flexible_2017}. With the VCE, contributors from remote areas can be recruited to perform virtual exploration, and teleported to the points where they are most needed, meaning that the mapping organisation does not have to be concerned about availability of workers, or optimising their budget based on the physical distance workers have to travel to complete their tasks. Furthermore, virtual exploration eliminates all privacy concerns about revealing contributors' locations. However, it must be noted that our approach can only be applied to a subset of SC tasks: those that can be done on a street-view image of the space, and is limited by how up-to-date images are.


In the context of urban auditing, several works have proposed the use of virtual spaces for expert auditors to carry out infrastructure verifications as an alternative to sending them to the field~\cite{badland_can_2010}. For example, FASTVIEW~\cite{griew_developing_2013} supports the exploration of nine street characteristics (e.g. pavement width and solidity) on top of Google Street View, while \citet{vanwolleghem_assessing_2014} studies the reliability and validity of a Google Street View-based safety assessment of cycling routes for children. These studies suggest that expert auditors are almost as accurate on a virtual space than on the field, potentially saving travelling costs. In our work, we attempt to prove that crowdworkers can be effective for the particular task of locating PoIs.






Other works used Google Street View to identify Street-level accessibility problems with crowdworkers~\cite{hara_combining_2013}. They selected a subset of flat images from Street View and developed an image labeling tool for crowdworkers from Mechanical Turk to signal accesibility problems, showing that untrained crowd workers were able to identify them. Further work by the same authors~\cite{hara_improving_2015} use Google Street View and crowdworkers to identify landmarks around bus stops, in order to create an accessibility map for low-vision or blind bus riders. Crowdworkers are teleported to the vicinity of a known bus location (based on the Google Transit API) and asked to label landmarks in or around it. Our work focuses on exploration tasks, where contributors need to move around the area of interest to locate PoIs instead of merely describing an item to where they were teleported. Our approach could be seen as complementary to this one, as it could be used to provide a map of locations that then could be audited in a separate task.
Recently, an open pilot of \emph{Project Sidewalk}~\cite{saha_pilot_2017} presented preliminary results of accessibility auditing in the style of \cite{hara_combining_2013} extended to an exploratory context, where contributors (paid or volunteering) are asked to inspect segments of street instead of being positioned in front of the particular location to audit. Their task is slightly different to ours in the sense that they require complex labelling of elements in a street (missing ramps, obstacles), while we focus on the location of a single type of item. As their work is preliminary, no analysis of completeness, accuracy or behavioural patterns was made available.




\section{System description and design}
\label{sec:systemdesign}

Our system aims to enable the recruitment of crowdworkers for generating maps of Points of Interests (PoIs) on top of street view imagery services. The end users  are mapping agencies and municipality departments in charge of urban infrastructure and mobility that need to locate PoIs within an area of interest. The definition of a PoI depends of the particular use case, organisations interested in completing their mobility infrastructure maps are interested in specific types of items (e.g., bike racks, disabled parking spots), whilst for organisations interested in urban auditing and compliance a PoI could be a potential hazard on the street, or areas without proper accessibility for people with disabilities. As such, we identified the following requirements:

\begin{enumerate}
\item Target PoIs that crowdworkers will be asked to locate need to be configurable by end users to accommodate different use cases
\item The area that crowdworkers will explore needs to be configurable, so end users can direct resources to where there are most needed
\item Crowdworkers' interface needs to allow them to accurately report the coordinates of the PoIs they found
\item The system needs to provide means for the validation of PoIs submitted by crowdworkers, and for their aggregation, \emph{i.e.}, identifying coordinates submitted by different workers that correspond to the same PoI 
\item The number of crowdworkers to be recruited needs to be configurable, to adapt to varying area sizes, budgets and required number of independent detections
\end{enumerate}

\begin{figure}
\centering
\includegraphics[width=0.7\linewidth]{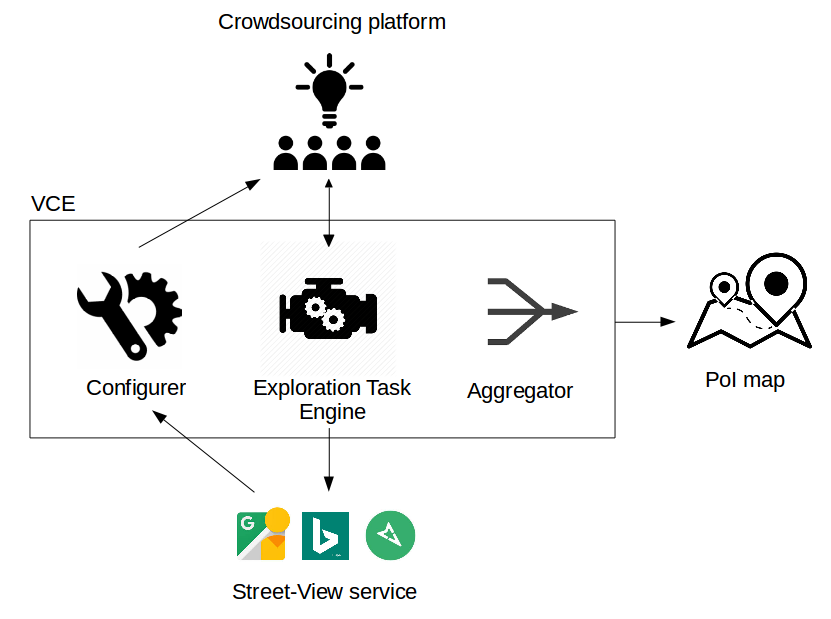}
\caption{High-level architecture of the Virtual City Explorer. The Street View service and the Crowdsourcing platform are assumed to be off-the-shelf. We developed the core of the VCE, comprised by the Configurer, Exploration Task Engine, and Aggregator.}
\label{figure:architecture}
\end{figure}

Based on these requirements, we defined a high-level architecture diagram of the system, shown in Figure~\ref{figure:architecture}. The first component (bottom of Figure~\ref{figure:architecture}) is a  service on top of which the other components rest. Companies like Google and Microsoft already offer such services, including APIs to embed imagery in 3rd party services and access metadata, meaning that they can be used off-the-shelf. For satisfying VCE requirements, a street view service needs to provide at least the coordinates that correspond to the street images, to allow workers to easily report PoI location, and the timestamp of the images, to assess the accuracy of the collected data with respect to the real world. Our implementation is based on Google Street View (GSV). The second component (Top of Figure~\ref{figure:architecture}) is a Crowdsourcing platform that enables the recruitment and payment of workers, and their redirection to the exploration task. Currently, several such platforms exist (FigureEight\footnote{\url{www.figure-eight.com}}, Mechanical Turk\footnote{\url{www.mturk.com}}, etc). Our implementation uses FigureEight. 

Finally, the VCE itself is comprised of the three components depicted in the center of figure~\ref{figure:architecture}. First, the \emph{Configurer} that enables end users to configure the parameters of the exploration task for workers: the area to be explored; the type of PoI that workers will be asked to locate; the number of workers to be recruited; and their starting positions. The output of the \emph{Configurer} is passed to the \emph{Exploration Task Engine}, that instantiate the parameters received from the \emph{Configurer} in a task for each incoming crowdworker. This component also takes care of validating the individual submissions of each worker. The output of the \emph{Exploration Task Engine} is a set of coordinates of PoIs identified by all workers, that is passed to the \emph{Aggregator} component, whose task is to identify which of the submitted coordinates correspond to the same PoI, and exclude outliers.

\subsection{Configurer}

The \emph{Configurer} is a Web interface that allows an end user to set the parameters of the exploration task to be executed. First, the end user is presented with an embedded Google Maps view where she can draw a closed polygon that represents the area to be explored. The polygon is saved as GeoJSON object to be passed to the \emph{Exploration Task Engine}. After selecting the area, the end-user defines the target type of PoI to be located by workers. The definition takes the form of the instructions to workers in the worker interface. Instructions are input as an HTML page containing the explanation to workers of the sought PoI. 
It is suggested instructions to be written in simple language, yet comprehensive enough to avoid misunderstanding of what is required, and contains example images of the target PoI taken from the underlying  service, and, if relevant, include at least one example of each sub-type of PoI. 

Next, the end user sets the number of times she wants the task to be executed, having the option of disallowing the same worker to repeat the task. We assume that the end user has enough credit in the crowdsourcing platform to pay the workers. Then, the end user chooses the function to assign the starting position of workers. In our current implementation, starting points are assigned randomly from points on streets. Street view services (and in particular GSV) include images of the inside of some buildings, the on-street check rules out workers starting from inside a building, losing time figuring out how to get out to the street. Subsequently, the end user sets the number of PoI instances that each worker will be asked to locate ($numInstances$), and the reward for completing the task ($Reward$). An exploration task instance is considered complete if the worker submits $numInstances$ locations. Finally, when the end user launches the exploration task, the \emph{Configurer} connects to the FigureEight API to create, configure and start the task in the platform, and generates the links towards the exploratory task. Each time a worker accepts the task in FigureEight, she is redirected to the Exploratory Task. The GeoJSON defining the area of interest, the instructions, the number of workers, and the list of starting points are passed to the \emph{Exploration Task Engine}.

\subsection{Exploration Task Engine}
\label{subsec:taskengine}

\begin{figure}
\centering
\includegraphics[width=\linewidth]{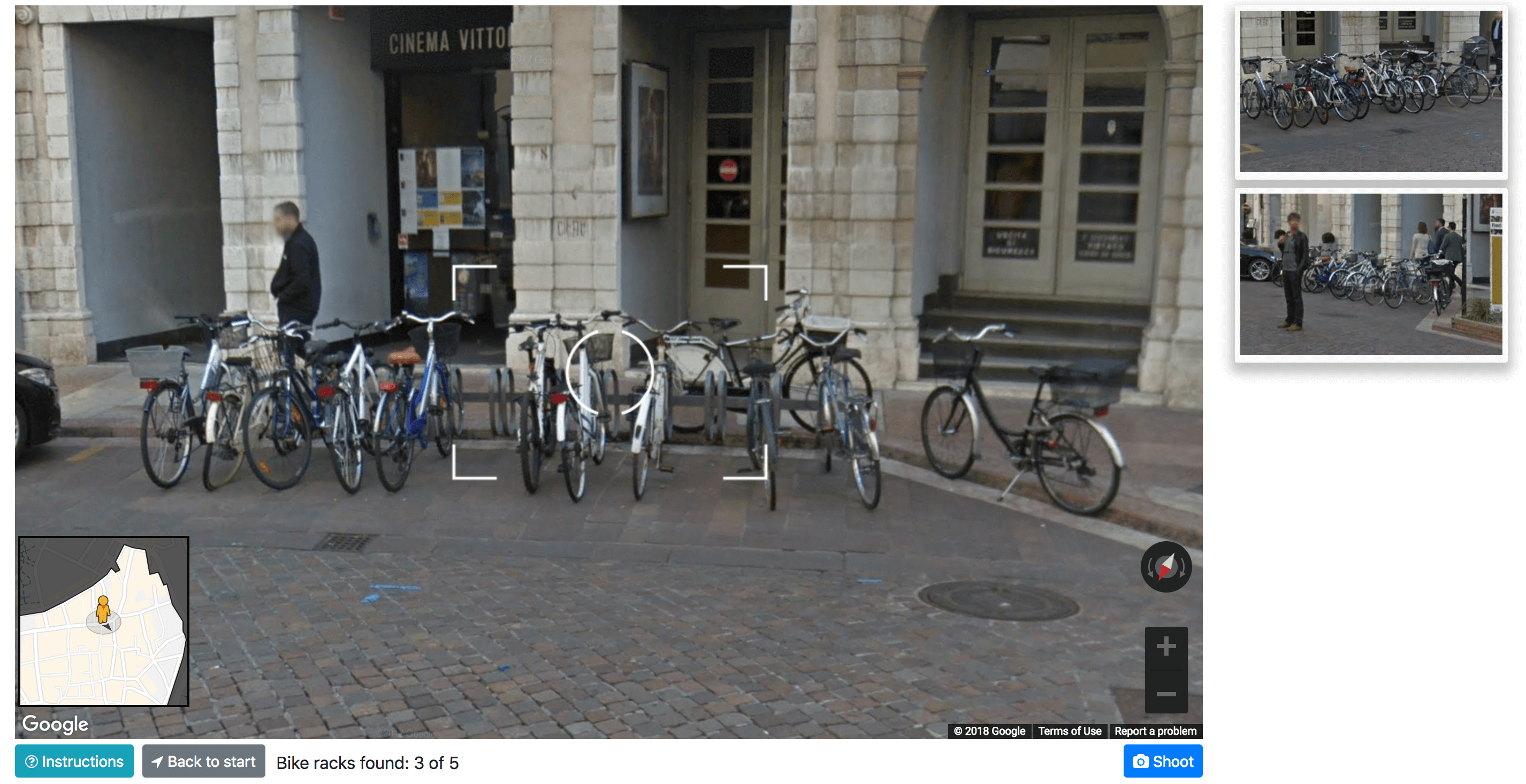}
\caption{Worker interface. The view finder in the middle combined with the Shot button (down-right) allows workers to take snapshots of PoIs from different angles (right side of the figure) and submit them for validation. Workers are informed of their relative position with respect to the area's boundary through the mini-map on the bottom left corner.}
\label{figure:microtask1}
\end{figure}

The \emph{Exploration Task Engine} is the central piece of the VCE. It implements the interface where workers explore the GSV imagery and report locations of PoIs, performs a first validation of the reported locations and stores intermediate results before aggregation. When a worker is redirected from the crowdsourcing platform, the engine instantiates a new exploration task within the configured area of interest, and assigns a starting point. The worker is first presented with the instructions set in the \emph{Configurer}. Once the worker has acknowledge them, it proceeds to the exploration interface, an \emph{HTML5} + \emph{JavaScript} page that connects to a \emph{MongoDB} database. Figure~\ref{figure:microtask1} shows the exploration interface from a worker point of view. The page contains an embedded GSV view that workers can explore using GSV controls: on-screen arrow buttons to move in the arrow's direction, or double-click on a distant point to ``fast-forward'' to it. To restrict exploration to the area of interest, whenever a worker attempts to navigate outside its boundary, she is automatically returned to her last location within the area, and a message explaining the reason for the repositioning is shown. Additionally, the interface shows the worker her relative position with respect to the boundary in a mini-map located on the bottom left corner. 

To allow workers to report PoI coordinates, we used the analogy of \emph{taking a picture}. Workers have available at all times a camera-style viewfinder and a \textbf{Shoot} button. When the shoot button is clicked, a snapshot of the current view is taken and shown at the right side of the screen, including metadata about worker's heading and coordinates. Workers are required to take three different pictures of each PoI, from three different angles. Workers can delete and update the pictures taken until they are satisfied. Once a worker has taken three pictures they are satisfied with, she can press the \textbf{Submit} button to submit the PoI location for validation.

\begin{figure}
\centering
\includegraphics[scale=0.35]{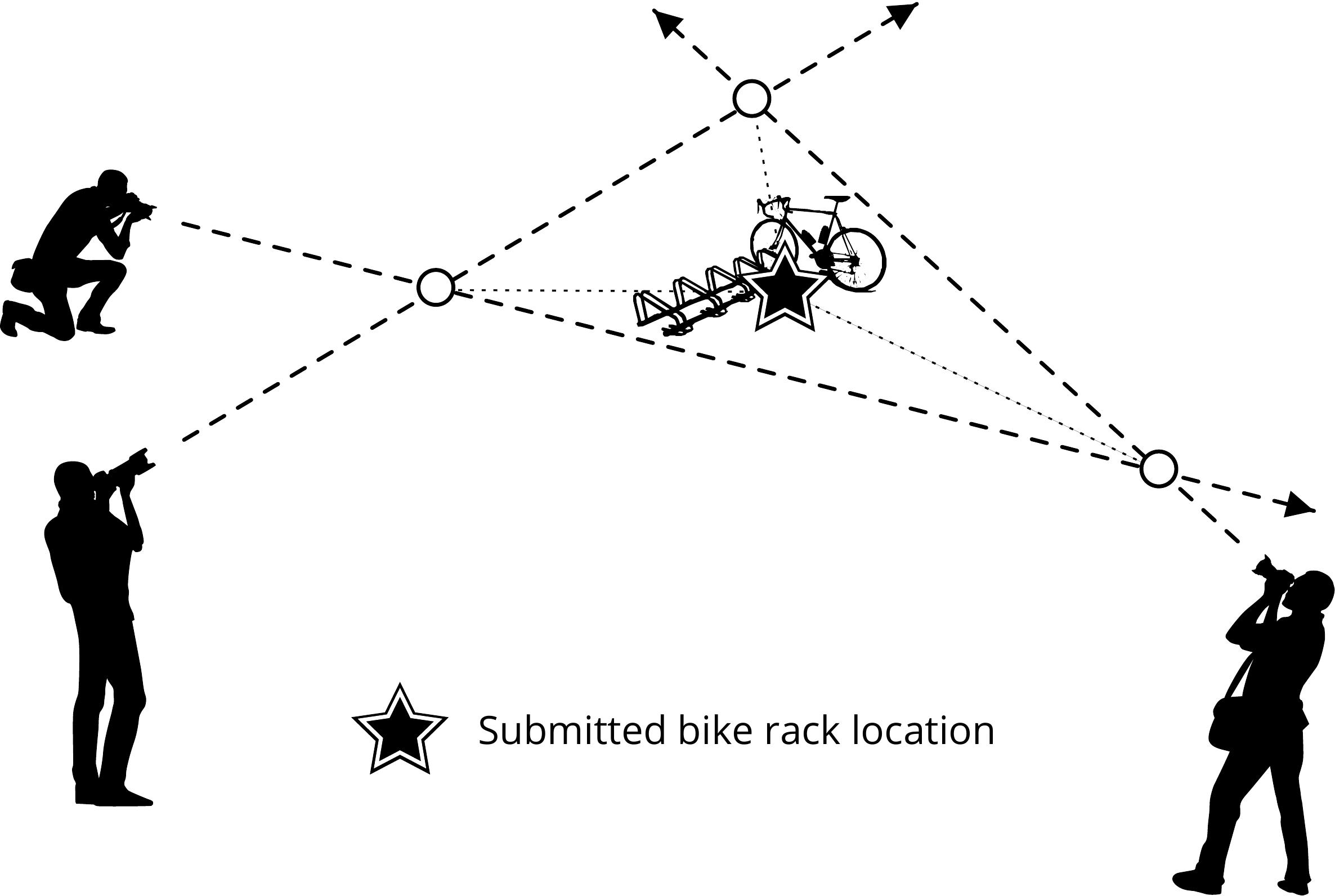}
\caption{Visual representation of the validation method. The centroid of the triangle formed by the intersection of the vectors of the pictures taken by workers  (here marked as a star) is reported as the (approximated) PoI location. A valid submission must have the centroid of the triangle at less than $\delta$ meters from all sides}
\label{figure:triangulation}
\end{figure}

When a worker submits a potential PoI, the engine verifies that the three pictures point to the same point in space, that is, to rule out submissions with three random pictures. To do so, it computes the intersection area of the vectors that start from the registered coordinates of the worker at each picture, as shown in Figure~\ref{figure:triangulation}. For a submission to be valid, the distance from the centroid of the triangle to each side needs to be less than a configurable distance (set to $10$ meters in our experiments). If the submission is valid, the coordinates of the centroid are registered as a PoI location. Otherwise, the submission is rejected and a message is shown explaining the reason.

Note that the engine does not check if the reported PoI is really an instance of the target PoI. Therefore, it is possible for a worker to report any object that passes the triangulation check and still get rewarded. The use of a crowdsourcing platform provides a first defence against this malicious behaviour, as when the culprit is eventually discovered, the end user can rate him negatively on the platform. In section
~\ref{sec:results}, we study the reasons for failure in connection with RQ1 in order to establish if workers behave honestly.

When a worker submits the configured $numInstances$ number of PoIs, the task finishes and she receives her reward throught the Crowdsourcing platform. Once the configured number of executions has been achieved, the task is unpublished and the collected list of coordinates is passed to the aggregator component. To enable the answering of our research questions our engine keeps an action log of worker's actions in the interface, including changes of position, pictures taken and discarded, and time spent in the task. However, the system deletes the action log and detected PoIs of workers that abandon the task before completing it based on ethical grounds: a dishonest end user may set too high parameters to make the task difficult to complete, and take advantage of partial results without having to reward crowdworkers. 

\subsection{Aggregator}
\label{subsec:aggregator}

Coordinates submitted by different crowdworkers need to be consolidated. The same PoI can be detected by different workers using different angles for their pictures, resulting in reported coordinates that are close to each other, but not exactly the same. The task of the aggregator component is to determine which points correspond to the same PoI. For this purpose, we use the DBSCAN algorithm~\cite{schubert_dbscan_2017} to cluster coordinates. DBSCAN requires two parameters: the minimum distance between points in a cluster and their neighbours ($Eps$), which we set to $10$ meters; and the minimum number of points within $Eps$ for the set to be considered a cluster ($MinPts$). 
Note that $MinPts$ can also be interpreted as the number of independent detections of the same PoI required to consider it \emph{confirmed} and include it in the final result. We defer to future work the study of variations of these parameters. We believe that $Eps$ should be set depending on the nature of the physical characteristics of the PoI, as wider ones (e.g. buildings) enable workers to report locations that are farther apart than smaller ones (e.g. trees).



\section{Experiments}
\label{sec:experimentdesign}

In this section, we describe the design and results of our experiments to answer research questions RQ1, RQ2, and RQ3 defined in section \ref{sec:intro}, to assess the feasibility, area coverage, and cost and time of the VCE, and compare them against VGI methods 

\subsection{Data}
\label{subsec:data}

The VCE can be adapted to find any type of PoI, but for this experiment we chose to fix the target to \emph{bike racks} for several reasons. First, there is a practical need for their location, due to the increasing interest in promoting more sustainable and green modes of transportation. Many bike racks are installed by private commercial property owners and municipalities cannot afford to send employees to explore all the city. From a crowdsourcing perspective, bike racks are challenging for workers due to their various shapes and sizes, and the potential confusion with other items where people attach their bikes (poles or fences), which also makes very hard their identification with machine learning classifiers.

As areas of interest, we chose two areas where bike rack maps collected with VGI methods or by experts: (i) the \emph{Zona a Traffico Limitato (ZTL)} - Limited Traffic Zone - of the city of Trento, Italy, which covers an area of 0.347 km$^2$, shaped as an irregular polygon. The city council sent one of its employees to collect bike rack locations within the ZTL, generating a bike rack map that is available as open data\footnote{\url{http://www.comune.trento.it/Aree-tematiche/Cartografia/Download/Rastrelliere-per-biciclette}}; (ii) a sector of the Penn Quarter neighbourhood of Arlington County in Washington D.C., USA. The department of Environmental Services of Arlington County developed a VGI mobile app called Rackspotter\footnote{\url{www.rackspotter.com}} to enable volunteers to contribute locations and metadata about bike racks, making the data openly available. The sector was chosen arbitrarily by us as a rectangle area of 0.33km$^2$, to include full segments of street and have a similar area than the ZTL. 

\begin{figure}
\centering
\includegraphics[width=0.49\linewidth]{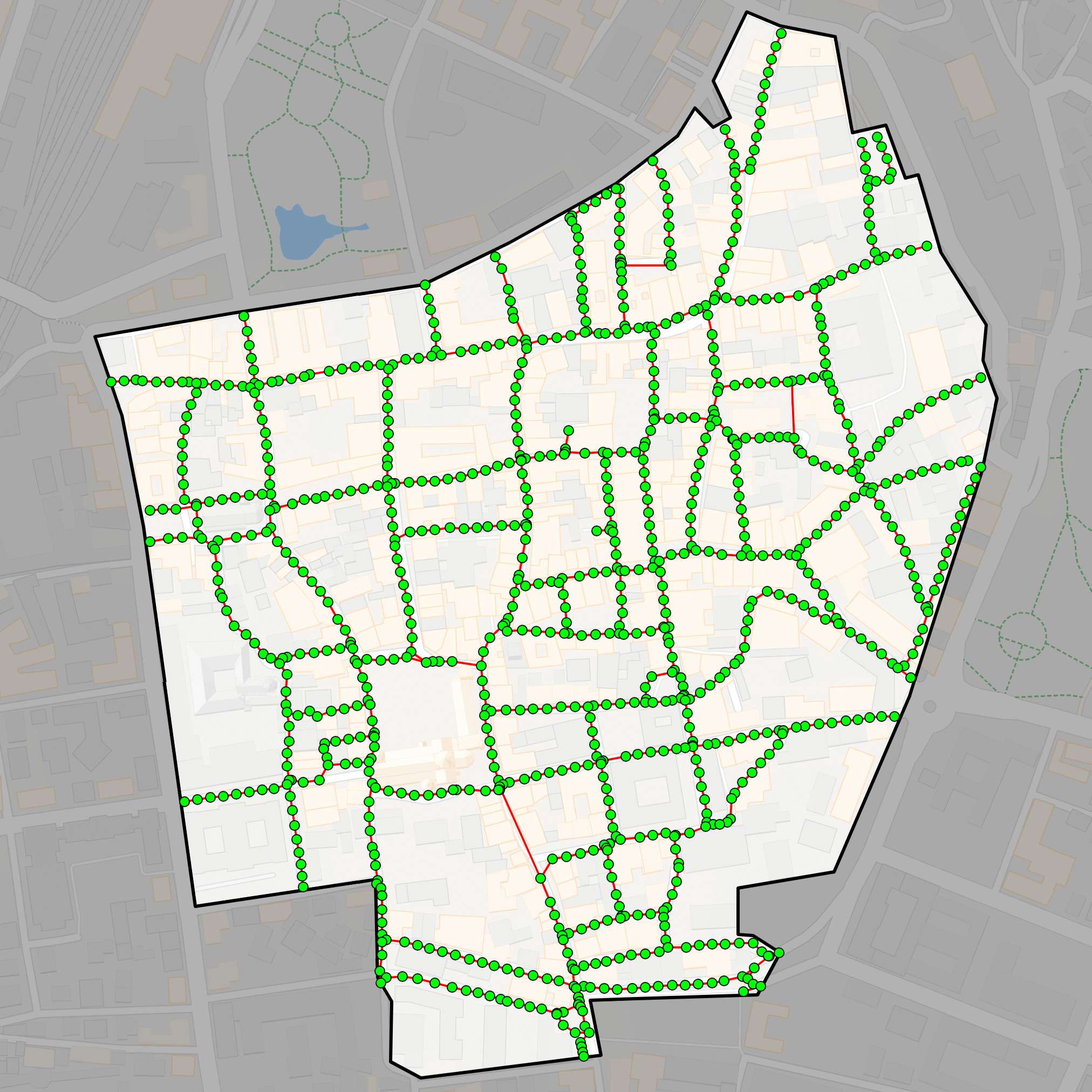}
\includegraphics[width=0.49\linewidth]{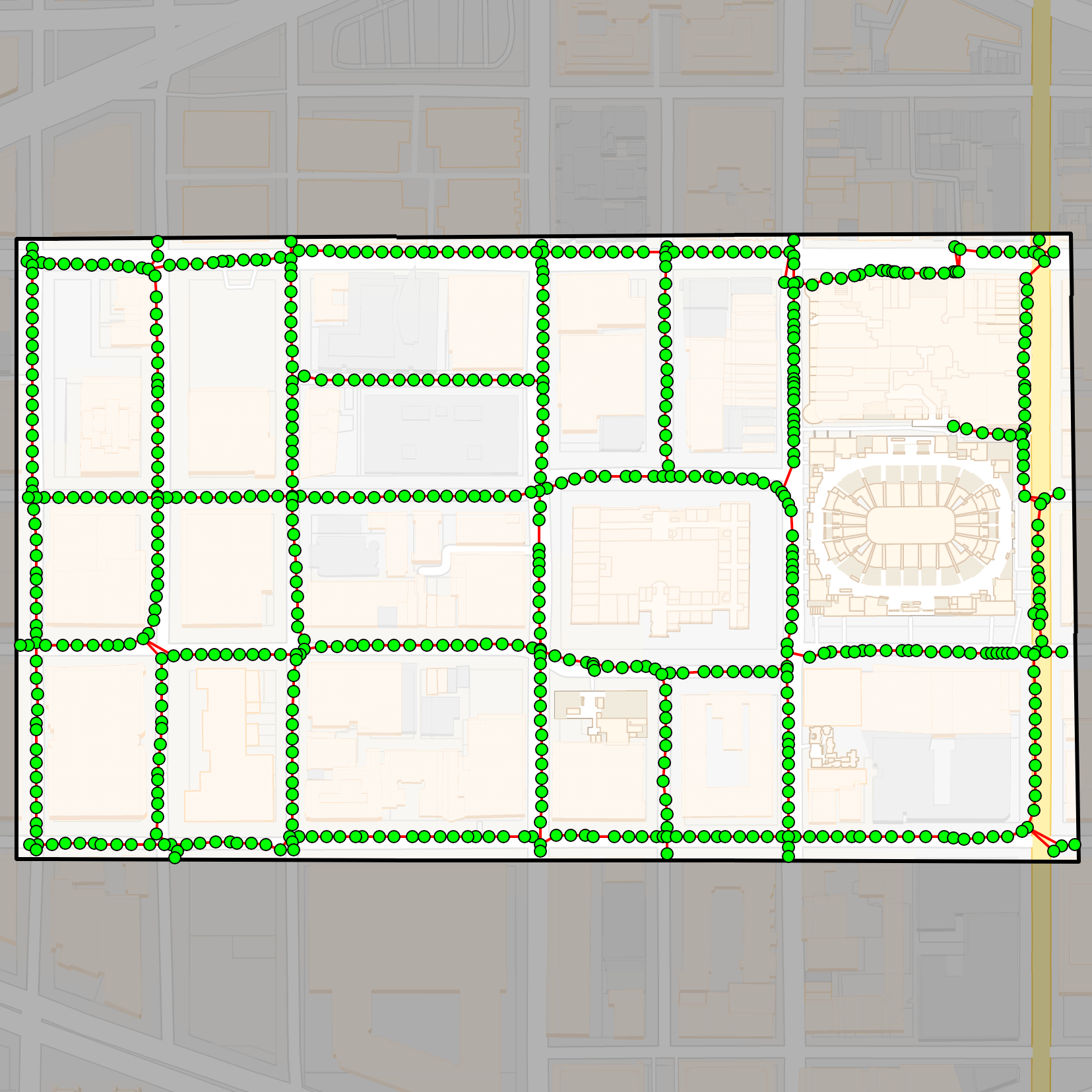}
\caption{Comparison of areas of interest: Trento's ZTL (left) and an area of Penn. Quarter, Washington D.C. (right). Green dots represent the explorable area, red lines, segments not covered by Google Street View.}
\label{figure:regions}
\end{figure}

As the coverage of street view services may not be complete in some areas, we define the \emph{explorable area} of an area of interest in a  service as the polyline where each point corresponds to a panoramic image centred in the coordinates of the point. We also define the \emph{explorable distance} as the length of the explorable area. We will use both definitions to compute the coverage of crowdworkers in the VCE (cf. RQ2). We computed the explorable areas of the ZTL and the Penn Quarter sector on Google Street View. Trento's ZTL explorable area has 924 points and 964 edges for a total explorable distance of $8869.38$m. Penn Quarter has 639 points and 653 edges for a total explorable distance of $6025.54$m. Figure~\ref{figure:regions} compares both areas side-by-side, with the green dotted line representing the explorable area, and red lines the not covered streets. We manually computed the length of the not covered streets: for Trento, $~605m$ ($6.5$ percent), and for Penn Quarter $~90m$ ($1.5$ percent).  Table \ref{tab:areas} summarizes the main characteristics of both areas.

\begin{table}[t]
\caption{Areas of interest used in the experiment}
\label{tab:areas}
\begin{tabular*}{\linewidth}{@{\extracolsep{\fill}}lcc}

\toprule
                     & \textbf{Trento - ZTL} & \textbf{Washington - Penn Quarter} \\ \midrule
Area                 & $0.347 km^{2}$ & $0.33  km^{2}$              \\ 
\# nodes in explorable area     & 924          & 639                       \\ 
\# segments in explorable area  & 964          & 653                       \\ 
Explorable distance  & 8869.3m      & 6025.54m  \\
Google Street View coverage & 93.5\%       & 98.5\%                    \\ 
\bottomrule
\end{tabular*}

\end{table}

 Table~\ref{tab:summary} compares the available bike rack maps for both areas of interest in terms of cost and time to generate, number of bike racks collected. Trento ZTL's bike rack map, \textbf{Municipality Map} in the table, was collected by sending one employee of their mobility unit with a dedicated GPS device, to explore the ZTL and report the location of bike racks. We privately contacted the Municipality to learn when the dataset was created and to get an estimation of time to generate and cost. The map was generated in half working day, with its cost estimated using the hourly salary of the employee that collected the data (\euro$15$). We also got confirmation that the map for the rest of the city could not be generated due to the higher cost of sending employees to explore larger areas.

Penn Quarter map, \textbf{Rackspotter} in the table, was collected through a VGI initiative\footnote{\url{http://rackspotter.com/}} run by the Transportation Demand Management agency of Arlington County, Virginia, USA. Volunteers use a mobile or a web app to report bike rack locations, either pinpointing on a map interface or the phone's GPS, and can submit additional information such as rack capacity, a photo, or other notes.  As  Rackspotter relies on volunteers, we assumed zero cost for generation. Unfortunately, we weren't able to obtain machine-readable data about contribution timestamps to compute the time taken to generate it or the most recent contribution. Also, as volunteers can contribute anonymously, it was not possible to know how many volunteers were involved in the process.

For both areas, we also compared with bike rack locations extracted from OpenStreetMaps, that we refer as \textbf{Trento OSM} and \textbf{Washington OSM} in the table. As with Rackspotter, we assumed zero cost of generation. We computed the time to generate as the difference between the timestamp of the first and last contributions in the area of interest. As OpenStreetMap contributors require an account, we were able to know that 2 volunteers made all contributions for Trento, and 4 volunteers did all contributions for Washington.




\subsection{Methodology}
\label{subsec:methodology}

Table~\ref{tab:parameters} summarizes the VCE parameters that we fixed for our experiments. As StartPoint function, we make each incoming crowd-worker start from a random point of the explorable area. A random starting point is easy to implement and could serve as a baseline for future experiments involving more elaborated functions. We generated one exploratory task for the ZTL and Penn Quarter areas of interest described in subsection~\ref{subsec:data}. We first published the Trento ZTL task, and after its $60$ executions were completed, we published the Washington task. Each worker was asked to locate five bike racks for each task execution, being paid $\$0.20$ after successful completion. We defer to future work the study of varying $numInstances$.

Regarding aggregation parameters, for $\delta$, we chose 10 meters, and as $Eps$ also 10 meters. Bike racks can be single slot small ones, but also wide to accommodate tens of bikes, making the decision of $\delta$ arbitrary, bike racks can also be relatively close to each other (e.g. on opposite sides of the same street), which led to our decision of using a relatively small $Eps$ value.  As $minPts$ we chose a value of 3.

\begin{table}[t]
\caption{VCE parameters and values set for experiments in this paper}
\label{tab:parameters}

\begin{tabular*}{\linewidth}{@{\extracolsep{\fill}}llr}
\toprule

\textbf{Parameter} & \textbf{Description} & \textbf{Value}               \\ 
\midrule
StartPoint & Function to assign starting points                    & Random \\ 
$\delta$      & Valid distance from triangle centroid & 10m                                   \\ 
$Eps$        & Distance to consider as same                            & 10m                                   \\ 
$Minpts$     & Cluster size for DBScan                                                   & 3                                     \\ 

$numExecutions$        & $\#$ of task executions (worker may repeat)                           & 60                                   \\ 
$numInstances$        & $\#$ of PoI instances each worker is asked to locate                           & 5                                   \\ 
$Reward$        & Reward after task completion                           & $\$0.20$                                   \\ 
\bottomrule
\end{tabular*}
\end{table}

The resulting bike rack maps were then analysed and compared with the OSM, Rackspotter, and Municipality maps as described in the following subsections.

\subsubsection{Method for RQ1. Feasibility and completeness}
\label{subsubsec:method-completeness}

To verify if crowdworkers using the VCE can accurately identify and locate bike racks, we assumed the OSM and Muncipality/Rackspotter maps as \emph{gold standards} and computed the following Information Retrieval metrics, assuming $A$ the map generated by our approach and $B$ a gold standard:

\textbf{The cardinality of the intersection} $(|A \cap B|)$ shows the number of bike racks in the gold standards also identified by VCE crowdworkers. 

\textbf{The cardinality of the set difference}  between the gold standard and our map $(|A - B|)$ shows the number of bike racks that crowdworkers using our system were not able to discover. We manually checked in Google Street View the coordinates of the missing bike racks and classified the failures as follows: (i) \emph{Absent}, when the bike rack was not at sight in the image; (ii) \emph{Unreachable}, when the bike rack was located inside a courtyard or building not accessible by Google Street View; (iii) \emph{Missed (Clustering)}, when crowdworkers identified a bike rack that indeed appears on Street View, but the coordinates computed by our approach placed it more than $10$ meters away from the coordinates in the gold standard; (iv) \emph{Missed}, when Google Street View showed a bike rack in the location specified by the gold standard, but our crowdworkers were not able to detect it; and (v) \emph{Triangulation impossible}, when the bike rack was in a spot where there were fewer than three images available on Google Street View, making impossible for workers to take the required three pictures to report it.  A large number of errors of type (i), (ii), and (v) would uncover a limitation of the use of Google Street View. A large number of errors of type (iv) would mean that crowdworkers are not effective finding bike racks. Errors of type (iii) need further analysis, as it is not possible to determine if both are the same bike rack, or if they are different and each approach correctly located one and missed the other.

 \textbf{The cardinality of the set difference} $(|B - A|)$ shows the number of possible false positives reported by our approach. We manually checked the pictures corresponding to false positives and classified them as: (i) \emph{Worker errors}, when they did not contain a bike rack; and \emph{Potential new bike racks}, when they did show a bike rack, meaning that the conflict with the gold standard is due either to the GSV image being outdated, \emph{i.e.}, the bike rack detected by crowdworkers does not exist any more, or due to the bike rack not being detected by the other method. To provide further insight, we compared the GSV timestamp of the image were they appeared with the timestamp of the latest contribution in the gold standard. A later timestamp in our approach would suggest that the bike rack was missed by the other method, while the converse would suggest that the bike rack was removed at a later date than the Street View image timestamp. A large number of type (i) errors would suggest that crowdworkers are not effective in the task. A large number of type (ii) errors would suggest that, either our approach is able to find bike racks not detected by the other approaches, or that the GSV image is too outdated, and many bike racks have been removed since. 
 
\textbf{The Jaccard similarity} ($J(A,B)$) is a metric of set similarity and diversity. We use it to quantify the overlap between bike rack maps. 

\subsubsection{Method for RQ2. Area Coverage}
\label{subsubsec:method-coverage}

To measure area coverage, we counted the number of points of the explorable area that were visited by crowdworkers  and divided it by the number of total points. We also logged the number of worker visits per point, and plotted a heatmap of each area to help uncover behavioural patterns.

\subsubsection{Method for RQ3. Cost and Time}

For our approach, we report the sum of the reward paid to workers for the $60$ executions. Completion time was computed as the timestamp difference between the time the task was made available in Crowdflower and the time the $60th$ execution was completed. For the \textbf{Municipality Map}, we privately contacted the Municipality of Trento to learn when the dataset was created and to get an estimation of the associated effort (half of a working day) and of the average salary of the mapper (\euro$15$). We also highlight the fact that the Trento municipality was interested in further information about our approach, as bike racks outside the ZTL have not being mapped due to lack of resources, and the fact that the bike racks in our map have 3 independent confirmations, while sending 2 or 3 different employees to independently confirm in-place is not possible for them. OpenStreetMaps have no cost besides the effort the volunteers invest for free. We computed the completion time as the timestamp difference between the first and the last contribution inside the area of interest. For \textbf{Rackspotter}, we assumed no cost as well, as we were uncertain if an investment on volunteer engagement has been carried out. Unfortunately, we could not obtain machine-readable data of contribution timestamps. 

\section{Results}
\label{sec:results}

\subsubsection{RQ1. Feasibility and Completeness}

Table~\ref{tab:Compare_Trento} (top) compares the maps generated with our approach and the reference datasets for the Trento area. The three methods generate quite different results, and there is no approach that clearly subsumes the others. 
Compared to the Municipality map, our approach is able to identify 27 out of 39 bike racks, missing 17 and detecting 5 potential new bike racks. If instead, we fix OSM as the gold standard, our approach is able to identify 27 ouf of 52 bike racks, missing 35 and detecting 8 potential new bike racks. We checked all potential new bike racks detected by our approach in GSV, finding that all of them corresponded to a bike rack, suggesting they were missed by the municipality employee and the OSM contributors. 


Table~\ref{tab:invest_trento_gt} summarises the reasons for failure regarding bike racks that were not detected by our approach in Trento. With respect to the Municipality map, all reasons are equally common. However, when compared to OSM map, non-reachability and absence from GSV are the main failure reasons. The latter reason might mean that some bike racks disappeared from the map since the last OSM update. 

Table~\ref{tab:Compare_Trento} (bottom) shows the comparison between the map generated with our approach and the quasi-gold standards for Penn Quarter. The OSM map has significantly less bike racks than the others (only 11); we believe this might be due to the county running an official VGI initiative that attracted more volunteers. As the OSM map is clearly less complete than the others, we focus our comparison on Rackspotter vs. our approach. Compared to Rackspotter, our approach was able to identify 37 out of 129 bike racks, missing 103 and identifying 11 potential new bike racks. This suggests that Rackspotter is more complete than our map.  We believe that the lack of completeness of our approach is due to our choice of 60 executions being insufficient for the bike rack \emph{density} in Washington. An alternative is to increase workers' goal to more than 5 bike racks. Research on which is the best strategy is deferred to future work.  We verified that all 11 potential bike racks were not false positives but unfortunately, due to the lack of timestamps for Rackspotter, we cannot confirm if they are indeed bike racks missed by volunteers, or if there is a chance they were added later.


\begin{table*}[t]
\caption{Pair-wise comparison of the overlap and similarity between bike rack maps}
\label{tab:Compare_Trento}
 \centering

\begin{tabular*}{\textwidth}{@{\extracolsep{\fill}}llccccccc}
\toprule
\textbf{Map A} &   \textbf{Map B} & \textbf{$|A|$} & $|B|$ & $|A \cup B|$ &     $|A \cap B|$ &       $|A-B|$ &       $|B-A|$  & $J(A,B)$  \\

\midrule

\textbf{Trento} \\

Municipality  & Crowd & 39 & 27 & 44 & 22 (50\%) & 17(39\%) & 5 (11\%) & 0.5 \\

OpenStreetMap & Crowd & 54 & 27 & 62 & 19 (31\%) & 35 (56\%) & 8 (13\%) & 0.31 \\
OpenStreetMap & Municipality & 54 & 39 & 66 & 27 (41\%) & 27 (41\%) & 12 (18\%) & 0.41 \\


\textbf{Washington} \\

Rackspotter   &        Crowd &  129 & 37 & 140 & 26 (19\%) & 103 (74\%) & 11 (8\%) & 0.19 \\
OpenStreetMap &        Crowd &  11 & 37 & 44 & 4 (9\%) & 7 (16\%) & 33 (75\%) & 0.09 \\
OpenStreetMap &  Rackspotter &   11 &  129 &   135 &    5 (3.7\%) &    6 (4.4\%) &  124 (91.9\%) & 0.04\\
\bottomrule
\end{tabular*}

\end{table*}


\begin{table}[t]

\caption{Failure reason analysis of bike racks not detected by our approach with respect to quasi gold standards} 
\begin{tabular*}{\linewidth}{@{\extracolsep{\fill}}lcccc}
\toprule
 & \multicolumn{2}{c}{ \textbf{Trento} } & \multicolumn{2}{c}{ \textbf{Washington} }\\
\textbf{Failure Reason} & \textbf{Municipality} & \textbf{OSM} & \textbf{Rackspotter} & \textbf{OSM}\\
\midrule
Missed & 11 (64.7\%) & 11 (31.4\%) & 92 (89.3\%) & 4 (57.1\%)\\
Absent from Street View & 2 (11.8\%) & 8 (22.9\%) & 4 (3.9\%) & 1 (14.3\%)\\
Unreachable & 2 (11.8\%) & 8 (22.9\%) & 2 (1.9\%) & 0 \\
Missed (clustering reasons) & 2 (11.8\%) & 4 (11.4\%) & 5 (4.8\%) & 2 (28.6\%) \\
Triangulation impossible & 0 & 4 (11.4\%) & 0 & 0\\
\midrule
All & 17 & 35 & 103 & 7\\
\bottomrule
\end{tabular*}

\label{tab:invest_trento_gt}
\end{table}








\subsubsection{RQ2. Area coverage}
\label{subsec:coverage}

\begin{figure}
\centering
\includegraphics[width=0.35\linewidth]{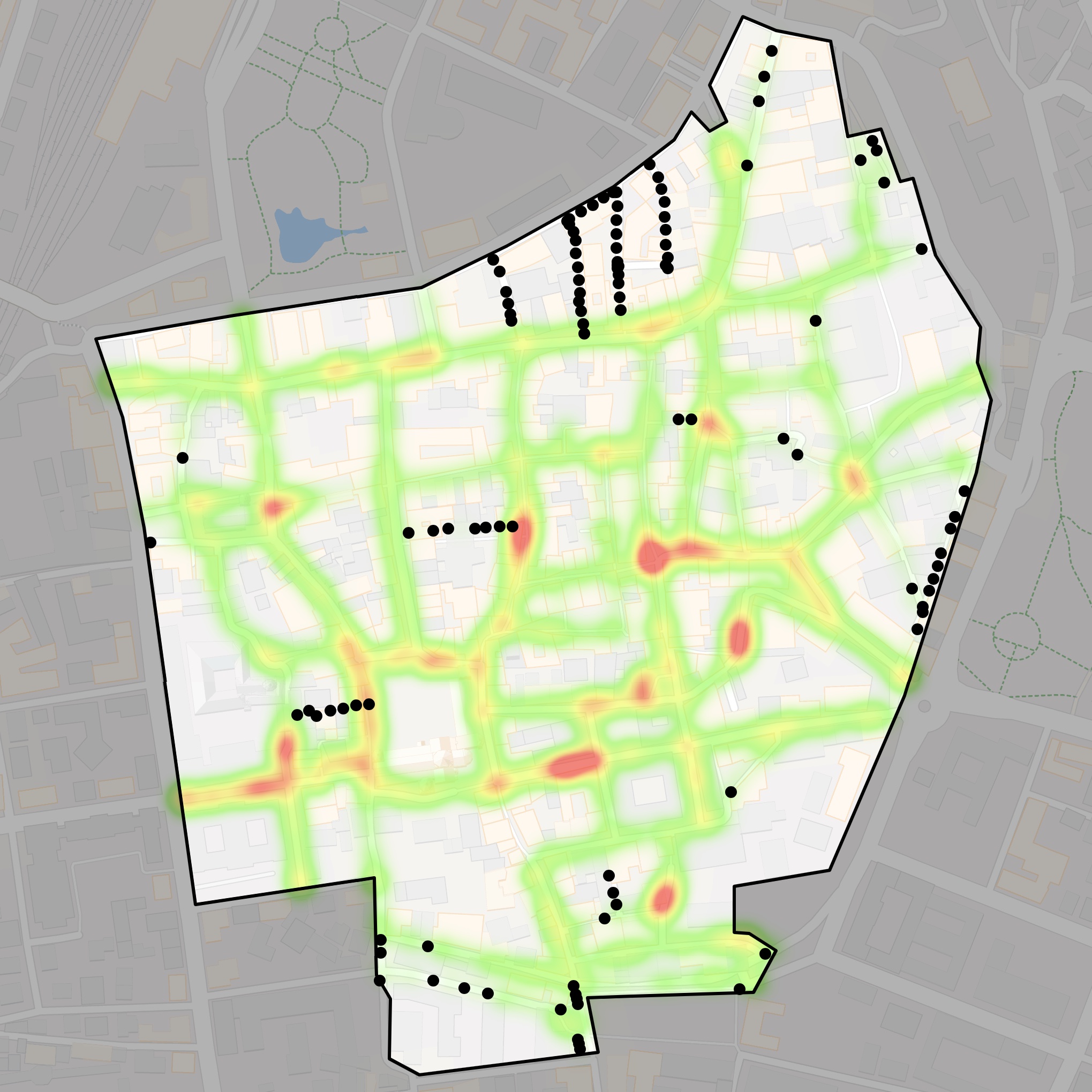}
\hspace{0.7cm}
\includegraphics[width=0.35\linewidth]{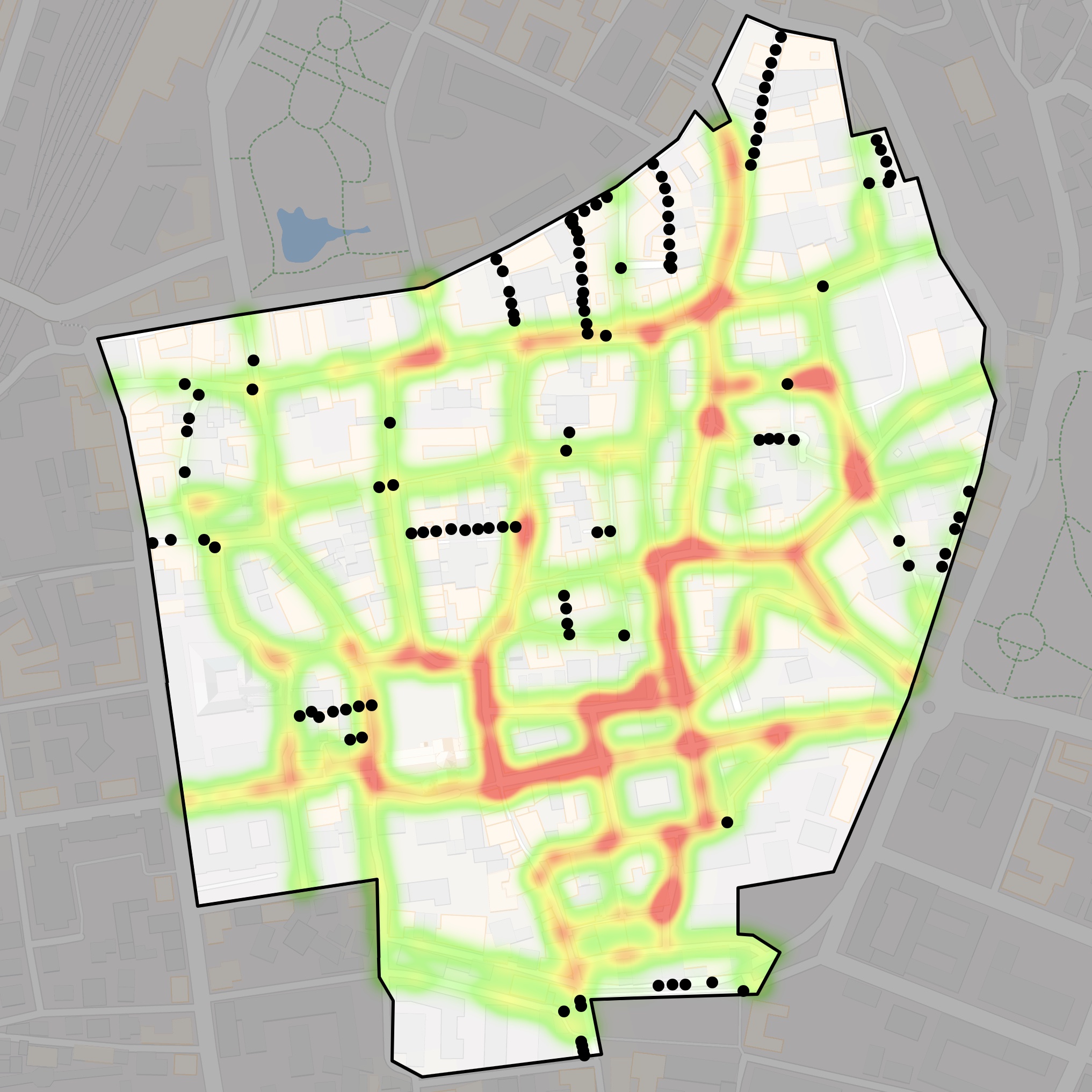}
\caption{Workers geographical coverage of basic (left) and taboo (right) task in Trento ZTL and Washington Penn}
\label{figure:heatmap_trento}
\end{figure}

\begin{figure}
\centering
\includegraphics[width=0.35\linewidth]{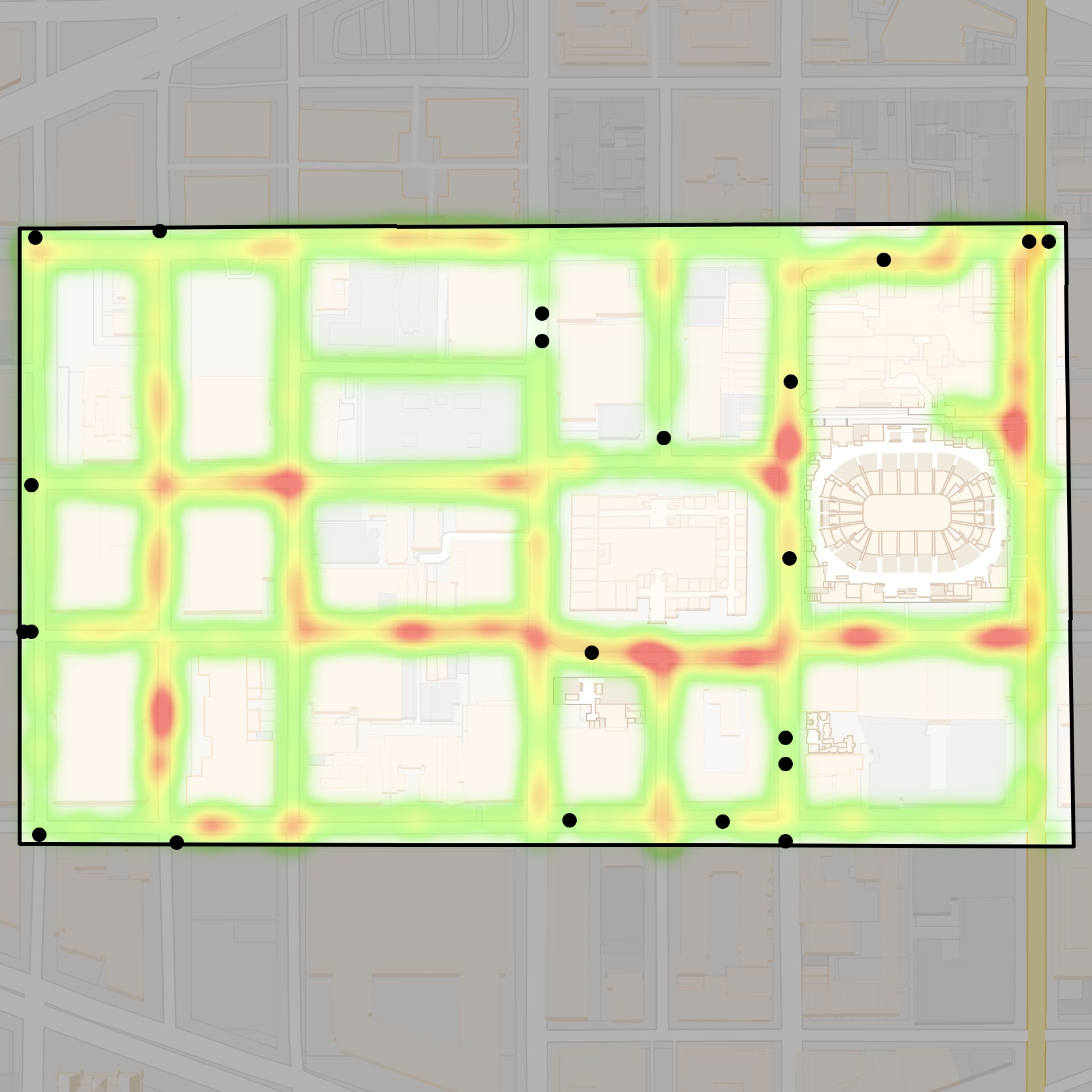}
\hspace{0.7cm}
\includegraphics[width=0.35\linewidth]{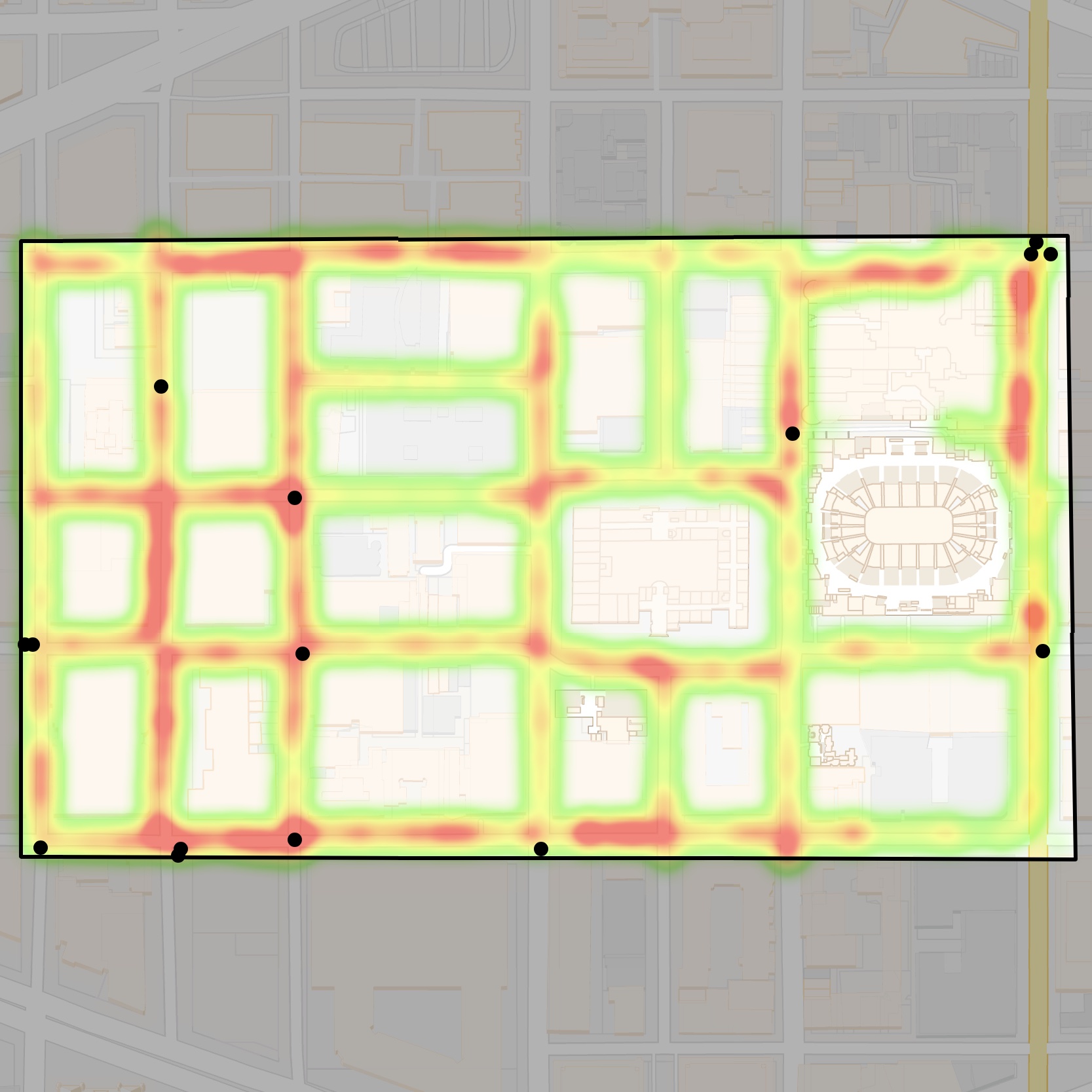}
\caption{Workers geographical coverage of basic (left) and taboo (right) task in Washington Penn.}
\label{figure:heatmap_washington}
\end{figure}

 For Trento, 88.2 percent of the points were visited at least once, while for Washington the ratio was 96.7 percent. Figure~\ref{figure:heatmap_trento} (left) shows a heatmap of worker's paths for Trento, and Figure~\ref{figure:heatmap_washington} (left) shows it for Washington. Black circles represent points that were not visited. In Trento, the frequency of visited points is lower near the border of the area, we also note that there are full segments of street that are not visited, especially on the north-east part of the region. 
For Washington, we first observed a highly visited point in the south-east corner of the area, further inspection revealed that one worker repeatedly tried to exit the area on that point (more than 300 times). We show the heatmap without that outlier. We also note that the not-visited points do not form a segment, meaning that all streets were visited.


\subsubsection{Cost and completion time}
\label{subsubsec:results-cost-time}

 Tables ~\ref{tab:summary} and \ref{tab:summary2} compares the bike rack maps available for each area against the ones generated with our approach in terms of costs; number of bike racks; time required to generate. For Trento, the municipality dataset was the fastest to generate, but at the highest cost. The cost of our approach is computed as follows: 60 workers times $\$0.2$ plus the FigureEight fee of $\$2.2$, for a total of $\$14.40$. 

\begin{figure}
\centering
\includegraphics[width=\linewidth]{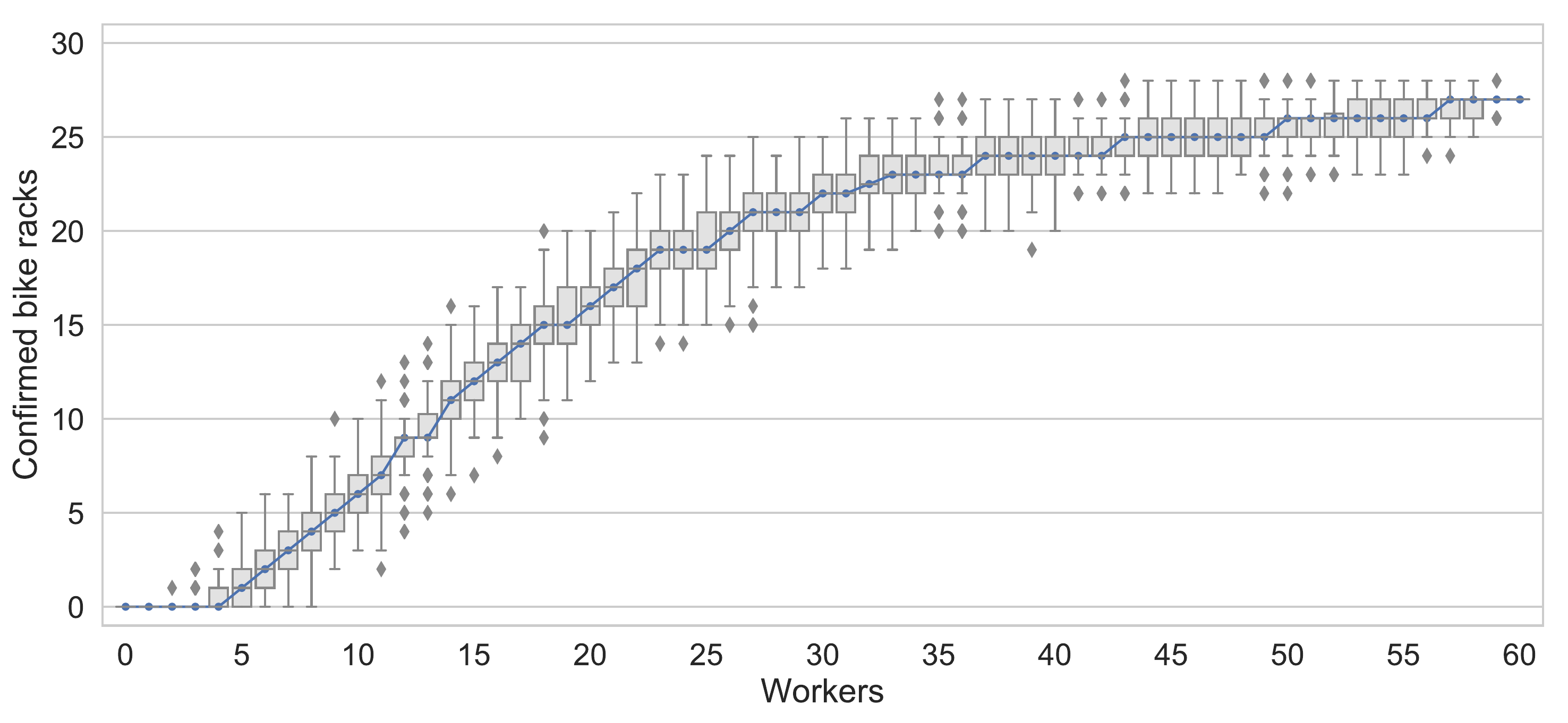}
\includegraphics[width=\linewidth]{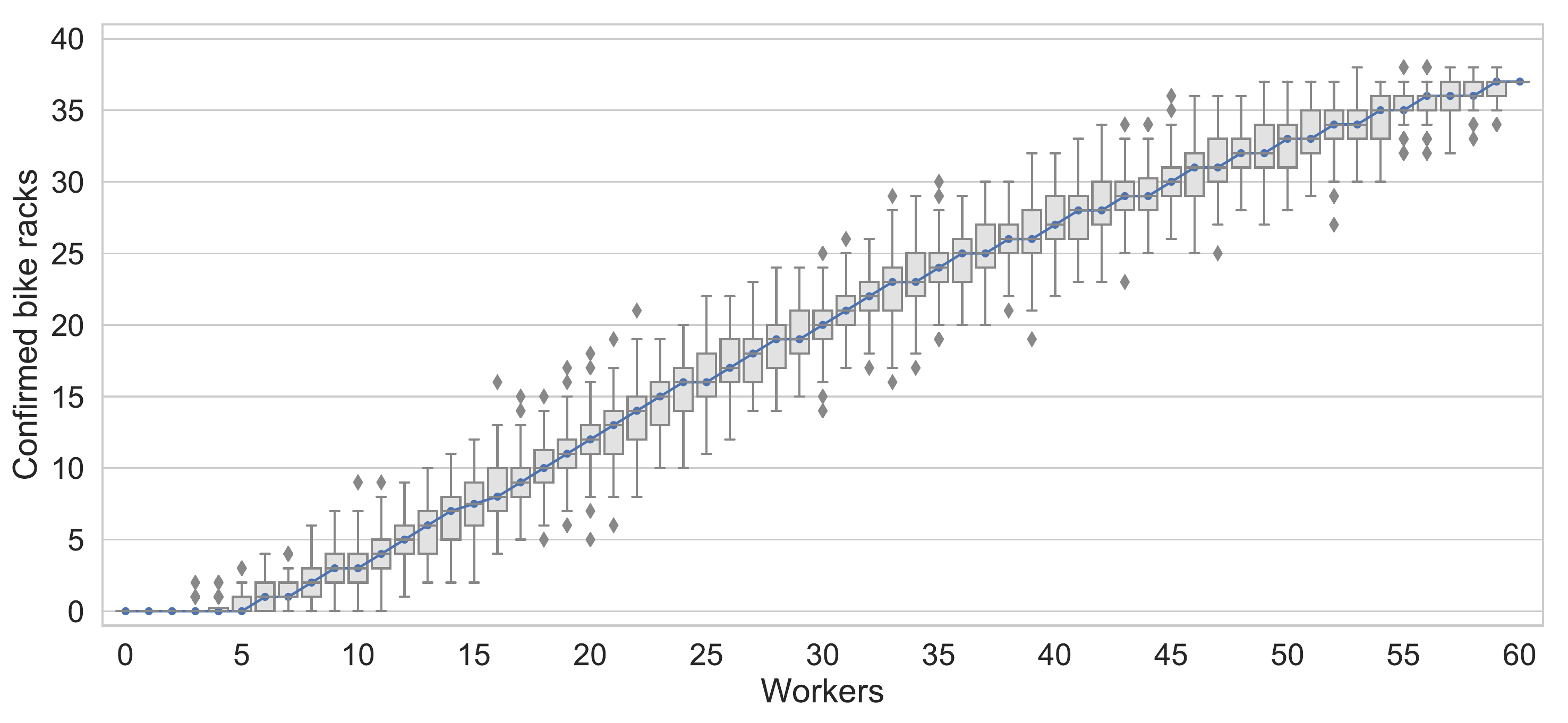}
\caption{Progression of number of confirmed bike racks over number of workers that complete the task for Trento (Top) and Washington (Bottom). Each value $n$  of the X axis was computed using 200 samples of size $n$ over the set of task executions}
\label{figure:workers_random_sampling}
\end{figure}

For the case of Washington, the OSM map was generated by four volunteers over a period of $74$ months, with the latest update from March $2016$; however, only $11$ bike racks were identified. This contrasts with the $52$ bike racks collected by the two volunteers that contributed to the Trento OSM map, and the $129$ identified by the Rackspotter community. Unfortunately, there is no machine-readable data available about the number of Rackspotter volunteers who contributed to the area of interest, or the timestamp of their contributions, to allow us to know if the difference between Rackspotter and OSM is due to Rackspotter having a larger or a more engaged community. We were also unaware if Rackspotter spends any significant amount of money for attracting volunteers. As the number of executions and goal for workers is the same as in the Trento case, the cost of our approach for Washington is also \$$14.40$. 
To understand how the number of task executions impacts our approach, therefore, its cost. From the set of 60 executions on each area, we sampled $200$ times a set of executions of size $n \in [0,60]$ and averaged the number of bike racks discovered in each set of samples.
Figure~\ref{figure:workers_random_sampling} shows the relation between number of task executions and number of reported bike racks. For the case of Trento, the relation resembles a logarithmic curve. We observe that around 35 workers were needed to obtain 20 confirmed bike racks, but from that point on, only 5 more confirmed bike racks were obtained with the remaining 25 workers. In the case of Washington, the relation is much more linear, there is no noticeable decrease on the slope until the 53rd worker, with the last 7 workers only contributing with two new confirmed bike racks.

We hypothesized that this convergence might be due to many workers submitting many times the same bike rack. To confirm this, we looked at the number of times each bike rack was independently detected, finding that $88\%$ of  Trento bike racks and $62\%$ of Washington bike racks were detected by 5 or more workers. Workers arriving late in the task submit bike racks already confirmed, wasting effort and preventing further bike racks to be discovered. To alleviate this issue, we developed the \emph{Taboo} strategy to guide incoming workers away from already confirmed bike racks.


\begin{table*}[t]
  \centering
    \caption{Overall comparison of bike rack maps generated with different approaches for Trento }
  
  \begin{tabular*}{\linewidth}{ @{\extracolsep{\fill}}lccc}
\toprule
 & \multicolumn{3}{c}{\textbf{Trento}}  \\

& \textbf{Municipality} & \textbf{OSM} & \textbf{Crowd} \\ 

\midrule

Cost& \euro60 & Free & \$14.4  \\

\# Bike racks& 39 & 52 & 27 \\

    Time to generate & Approx. 4 hours & 62 months & Approx. 3 days  \\
    
    
Most recent data & Jun. 2017 & Oct. 2015 & Oct. 2017  \\ 
    
Collected by & 1 expert & 2 volunteers & 60 crowdworkers  \\

\bottomrule
    
  \end{tabular*} 

  \label{tab:summary}
\end{table*}

\begin{table*}[t]
  \centering
    \caption{Overall comparison of bike rack maps generated with different approaches for Washington }
  
  \begin{tabular*}{\linewidth}{ @{\extracolsep{\fill}}lccc}
 
 \toprule
 
  & \multicolumn{3}{c}{\textbf{Washington}}  \\

&  \textbf{Rackspotter} & \textbf{OSM} & \textbf{Crowd} \\ 
\midrule

Cost&  Free & Free & \$10.80 \\

\# Bike racks&  129 & 11 & 37 \\

    Time to generate  & Unknown & 74 months & Approx. 3 days \\
    
    
Most recent data &  Unknown & Mar. 2016 & Jul. 2017 \\ 
    
Collected by &  Volunteers & 4 volunteers & 60 crowdworkers \\ 
    
    \bottomrule
  \end{tabular*} 

  \label{tab:summary2}
\end{table*}

\section{Taboo optimisation}
\label{sec:taboo}

In section \ref{subsubsec:results-cost-time}, we observed that a large number of bike racks were detected by more than 5 workers ($88\%$ for Trento and $62\%$ for Washington). This represents a waste of workers time and end user's resources. If a subset of PoIs is located in an easy to access area, workers will report those and neglect the exploration of further regions of the area. To avoid this, we extended our Exploration Task Engine with a \emph{Taboo} optimisation that works as follows: each time a PoI is identified by a predetermined number of independent workers, it becomes \emph{taboo}. Any subsequent worker starting the task will not be able to submit it, being shown an error message if she attempts to do so. To warn workers that the PoI is no longer selectable, a \emph{Taboo} red sign is shown on top of each PoI that has become taboo. However, the taboo optimisation raises a fairness issue: workers that arrive later in the task may find that all PoIs have already become taboo, forcing them to spend more time and effort to complete the task than others, or even abandon it without any reward. To counter this problem, we introduced a \emph{escape condition} based on two parameters: a combination of the time spent in the task and the distance covered: when a worker walks a predetermined distance \emph{and} has spent a predetermined time in the task without successfully detecting any PoI, the task finishes and the worker is rewarded as if she has submitted all required instances. Of course, care has to be taken to select the escape condition parameters: if too many workers escape without detecting any PoIs, the optimisation becomes useless.

To test if the \emph{taboo} optimisation improves the area coverage and completeness of the VCE (our RQ4), we generated two exploratory tasks with the taboo optimisation activated, using the same PoI (bike racks) as before, the same areas of interest described in section~\ref{subsec:data}, and the same parameters summarized in table~\ref{tab:parameters}. Table \ref{tab:tabooparameters} describes the parameters exclusive to the taboo optimisation and their values for our experimentation. We chose $TabooThreshold = 3$ to match the $Minpts$ value on the experimentation. The values for $escapeDistance$ and $escapeTime$ were chosen arbitrarily by us, we defer to future work the study of different escape conditions. 

The resulting maps were compared against the maps generated without the optimisation (that we call \emph{Basic} from now on). In terms of completeness, we used the same method described in section~\ref{subsubsec:method-completeness}: we assumed the map produced with Basic as gold standard and computed our battery of Information Retrieval metrics for the \emph{Taboo} maps. Furthermore, we compared the number of single \emph{detections} (i.e. every individual report of a PoI location) for each strategy, and the number of \emph{confirmed detections} (i.e. the bike racks on the final result). The number of single detections for Basic is always equal to $numExecutions \times numInstances$, if for Taboo is the same number, this would mean that the escape condition was never activated, and all workers were able to found $numInstances$ bike racks before triggering it. On the other hand, if some workers trigger the escape condition, the number of single detections for Taboo is less than for basic, and we are intuitively \'' saving" the effort of those detections. Of course, this saving only makes sense if we get the same number or more of confirmed detections, and the same or greater coverage. For measuring the impact of Taboo on area coverage, we use the same method described in section~\ref{subsubsec:method-coverage}: we compute the percentage of explorable area covered, and compared the heatmaps of Taboo against Basic. Finally, to verify if workers under the Taboo strategy explore more, we counted the number of virtual steps workers made after detecting a bike rack instances. We expect to see a higher number of steps for workers under Taboo.

\begin{table}[tbp]
\caption{Parameters of the Taboo strategy}
\label{tab:tabooparameters}

\begin{tabular*}{\linewidth}{@{\extracolsep{\fill}}lcc}
\toprule

\textbf{Parameter}  & \textbf{Description} & \textbf{Value}             \\ 
\midrule
TabooThreshold & \# of detections for making a PoI taboo                    & 3 \\ 
escapeDistance   & Distance walked to escape  & 1800m                                   \\ 
escapeTime     & Time without detecting a PoI to escape                                                  & 3 minutes                                     \\ 
\bottomrule
\end{tabular*}
\end{table}

\subsection{Results on completeness}
\label{subsec:taboo_completeness}

Table~\ref{tab:Basic_vs_Taboo} shows the pairwise comparison for each area of the maps generated for each strategy, similar to the comparison between the Basic strategy and the other maps done in section \ref{sec:experimentdesign}. We observe that for the case of Trento, $30\%$ more bike racks were confirmed and that except for one bike rack, the Taboo map is a superset of the Basic map. For the case of Washington, only three more bike racks were detected, and the relatively low Jaccard similarity value indicates that there is no much overlap among them. We believe this is a consequence of the high density of bike racks in the Washington area (129 as for Rackspotter) giving workers many options to select in a close area. Further experiments with PoIs with varying density are part of future work to test this hypothesis.

\begin{figure}
\centering
\includegraphics[width=\linewidth]{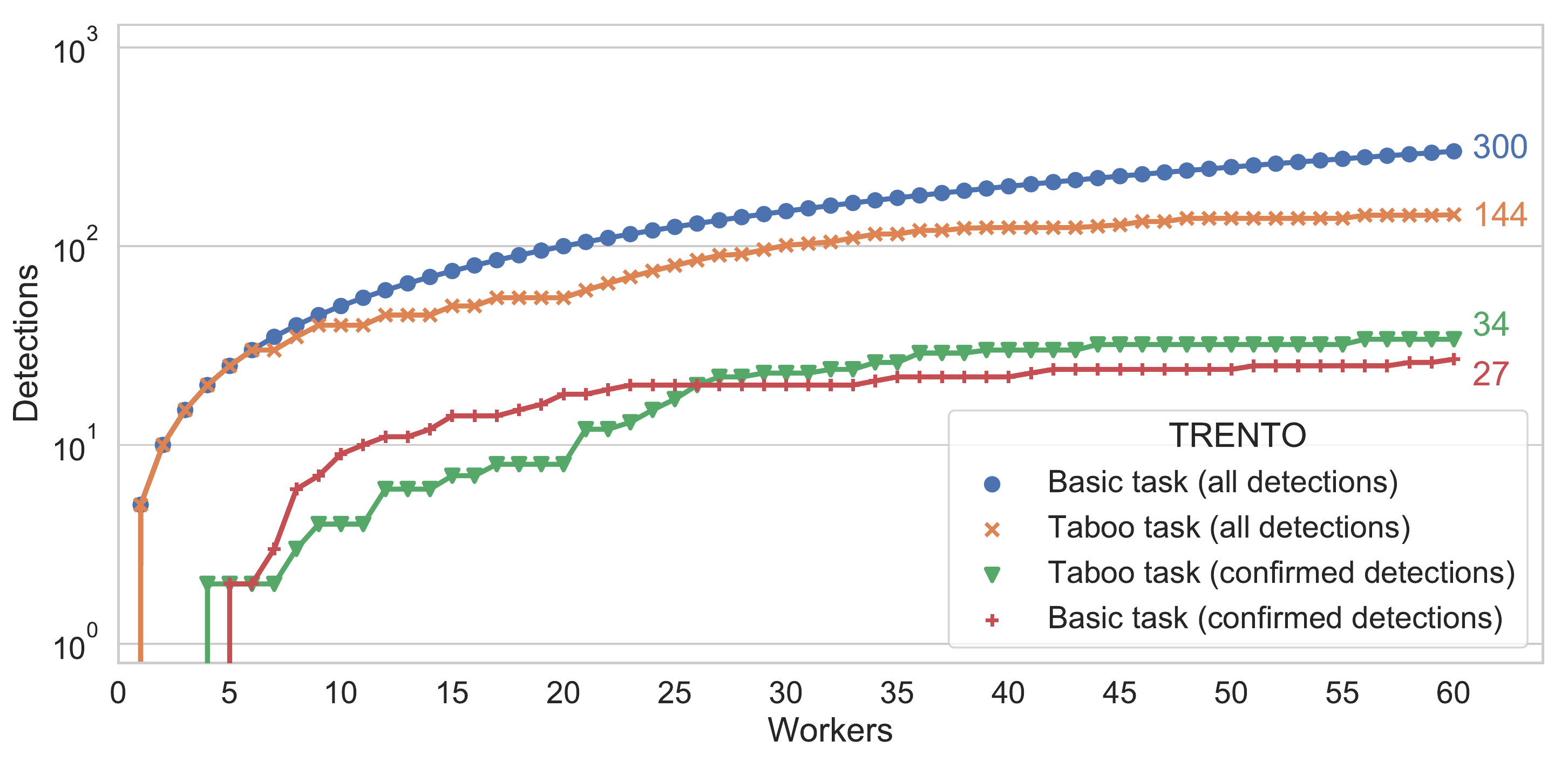}
\caption{Comparison of single and confirmed detections between Basic and Taboo strategies for the Trento area}
\label{fig:trento-basic_VS_Taboo}
\end{figure}

\begin{figure}
\centering
\includegraphics[width=\linewidth]{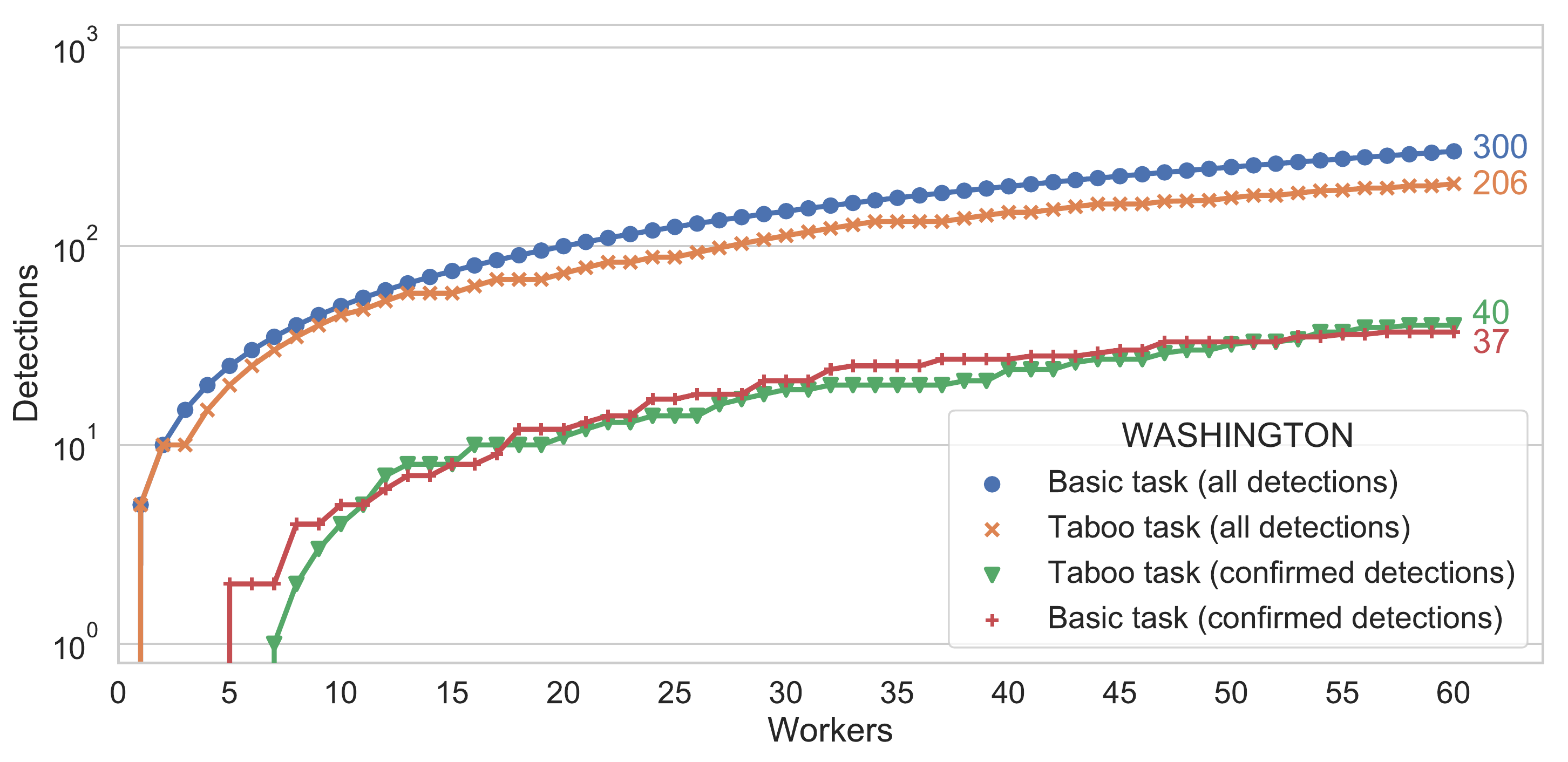}
\caption{Comparison of single and confirmed detections between Basic and Taboo strategies for the Washington area}
\label{fig:washington-basic_VS_Taboo}
\end{figure}

Figures \ref{fig:trento-basic_VS_Taboo} and \ref{fig:washington-basic_VS_Taboo} compare, for Trento and Washington respectively, the number of detections with respect to the number of workers that completed the task, ordered by timestamp of task completion in each experiment\footnote{We cannot apply a sampling method like we did for Figure \ref{figure:workers_random_sampling} because the order in which workers join the task  in the Taboo strategy impacts their selection, while in Basic, it is independent}, in log scale. The top two lines (blue and orange) compare the number of total bike rack detections for the Basic and Taboo strategies.  We observe that in the case of Trento, less than a half of the detections of the basic strategy, while in the case of Washington, is a little more than two thirds. The bottom two lines (green and red) compare the number of confirmed bike racks (detected by at least 3 different workers).  Note that after the 25th worker, the number of confirmed bike racks for the Basic strategy stalls, meaning that most detections are on already confirmed bike racks\footnote{and consistent with Figure \ref{figure:workers_random_sampling}}, while the Taboo strategy keeps generating new confirmed bike racks. Contrary to the case of Trento, we do not observe any apparent dominance of the Taboo strategy until the 55th worker.

\begin{table*}[t]
\centering
\caption{Pair-wise comparison of the overlap and similarity between bike rack maps between Basic and Taboo strategies}
\label{tab:Basic_vs_Taboo}
\begin{tabular*}{\textwidth}{ @{\extracolsep{\fill}} llccccccc }
\toprule
\textbf{Map A} &   \textbf{Map B} & \textbf{$|A|$} & $|B|$ & $|A \cup B|$ &     $|A \cap B|$ &       $|A-B|$ &       $|B-A|$ & $J(A,B)$  \\
\midrule
Trento Basic   &        Trento Taboo & 27 & 34 & 35 & 26 (74\%) & 1 (3\%) & 8 (23\%) & 0.74 \\
Washington Basic &        Washington Taboo &  37 & 40 & 50 & 27 (54\%) & 10 (20\%) & 13 (26\%) & 0.54 \\

\bottomrule
\end{tabular*}

\end{table*}

\begin{figure}
\includegraphics[width=\linewidth]{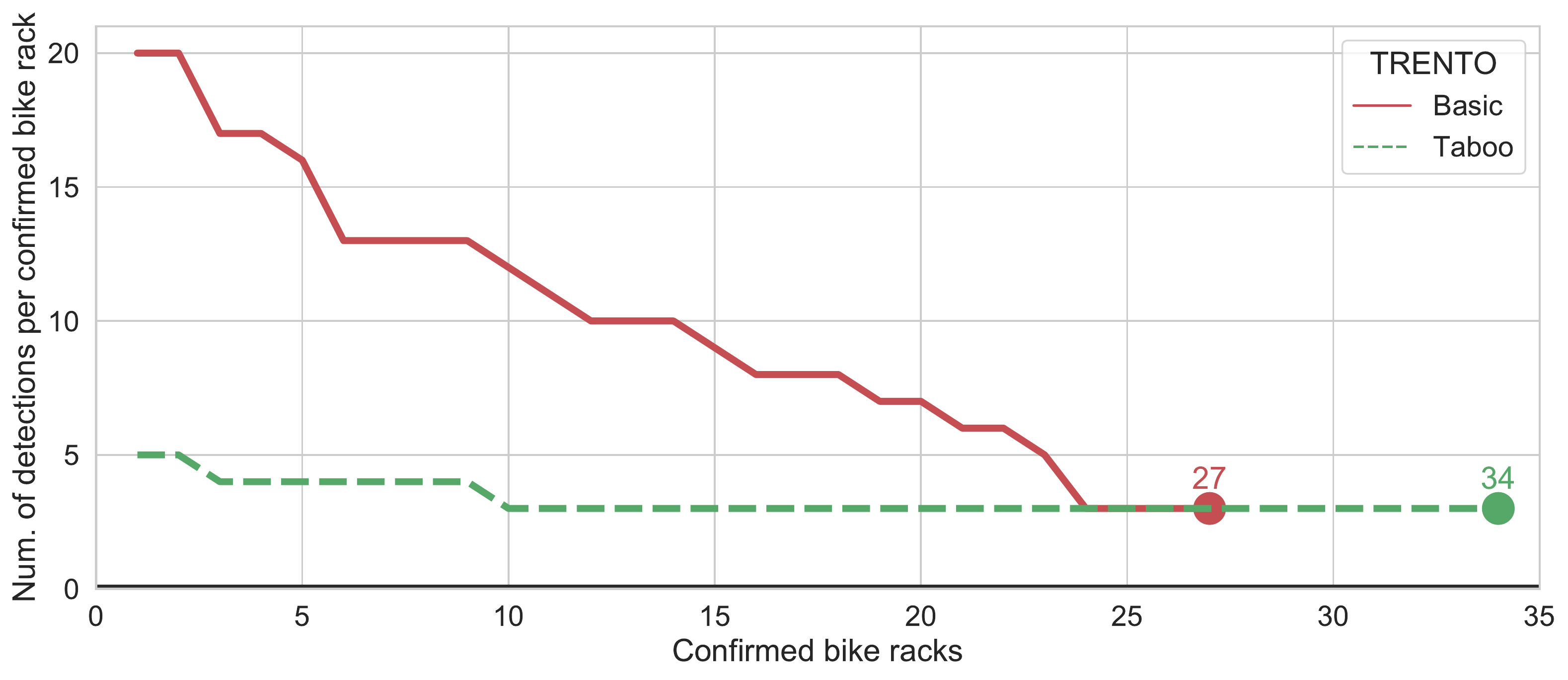}
\caption{Number of detections received per confirmed bike racks in Trento experiments for Basic and Taboo strategy.}
\label{fig:trento:impact_costs}
\end{figure}

\begin{figure}
\includegraphics[width=\linewidth]{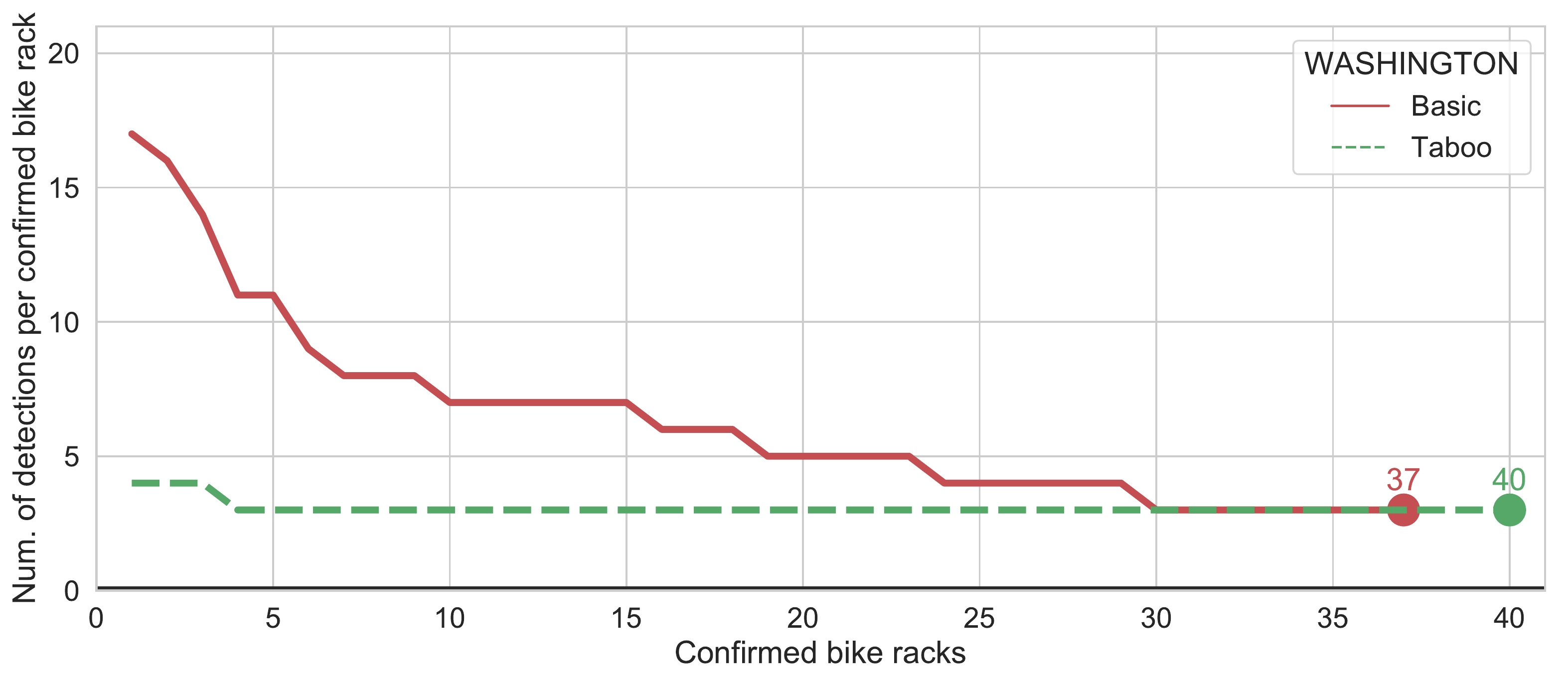}
\caption{Number of detections received per confirmed bike racks in Washington experiments for Basic and Taboo strategy.}
\label{fig:washington:impact_costs}
\end{figure}

To quantify the reduction of number of detections per confirmed bike rack. Figures \ref{fig:trento:impact_costs} and \ref{fig:washington:impact_costs} show for Trento and Washington respectively, the number of detections per confirmed bike rack for the Basic strategy (red solid line) and the Taboo strategy (green dashed line). By construction, the Taboo strategy should have at most 3 detections per bike rack, however, nearly 10 bike racks have 4 or 5 detections due to some workers exploring concurrently, and reporting the same bike rack (as the taboo list is only updated after a worker finishes, and applies to the next incoming worker). The area between the two lines quantifies the number of unnecessary detections saved.  

Our results suggest that the Taboo strategy does reduce the number of unnecesary confirmations, and increases the number of confirmed items, despite the fact that workers contribute with less detections due to the escape condition. We hypothesize that the lesser positive effect on the Washington area in terms of confirmed bike racks is due to the higher density of bike racks: when a bike rack becomes taboo, there are many different ones close enough to allow workers to complete the task before triggering the escape condition. Further experiments with PoIs with varying density are part of our future work to confirm this hypothesis.

\subsection{Results on area coverage}
\label{subsec:taboo_coverage}

Trento's coverage with the taboo strategy was  $87.2\%$, marginally less than the coverage achieved with the basic strategy reported in section \ref{subsec:coverage}. For Penn Quarter, the taboo coverage was $97.65\%$, practically equal to the coverage observed for the basic strategy. Figures \ref{figure:heatmap_trento} and \ref{figure:heatmap_washington} compare the heatmaps of the workers' paths for both strategies (left for Basic, right for Taboo). In the case of Washington, workers explored more the western side of the area. As it was the case of the Basic strategy, only a few isolated points were not visited. In the case of Trento with the Taboo strategy, more workers explore an area close to the center of the polygon, but they left unexplored almost the same segments of street that were unexplored with the Basic strategy.  We believe these segments have either concealed entrances or it is not immediately apparent that one can go through them, motivating future work on dynamic starting point functions that place incoming workers in areas detected to be not covered. 

\begin{figure}
\centering
\includegraphics[width=\linewidth]{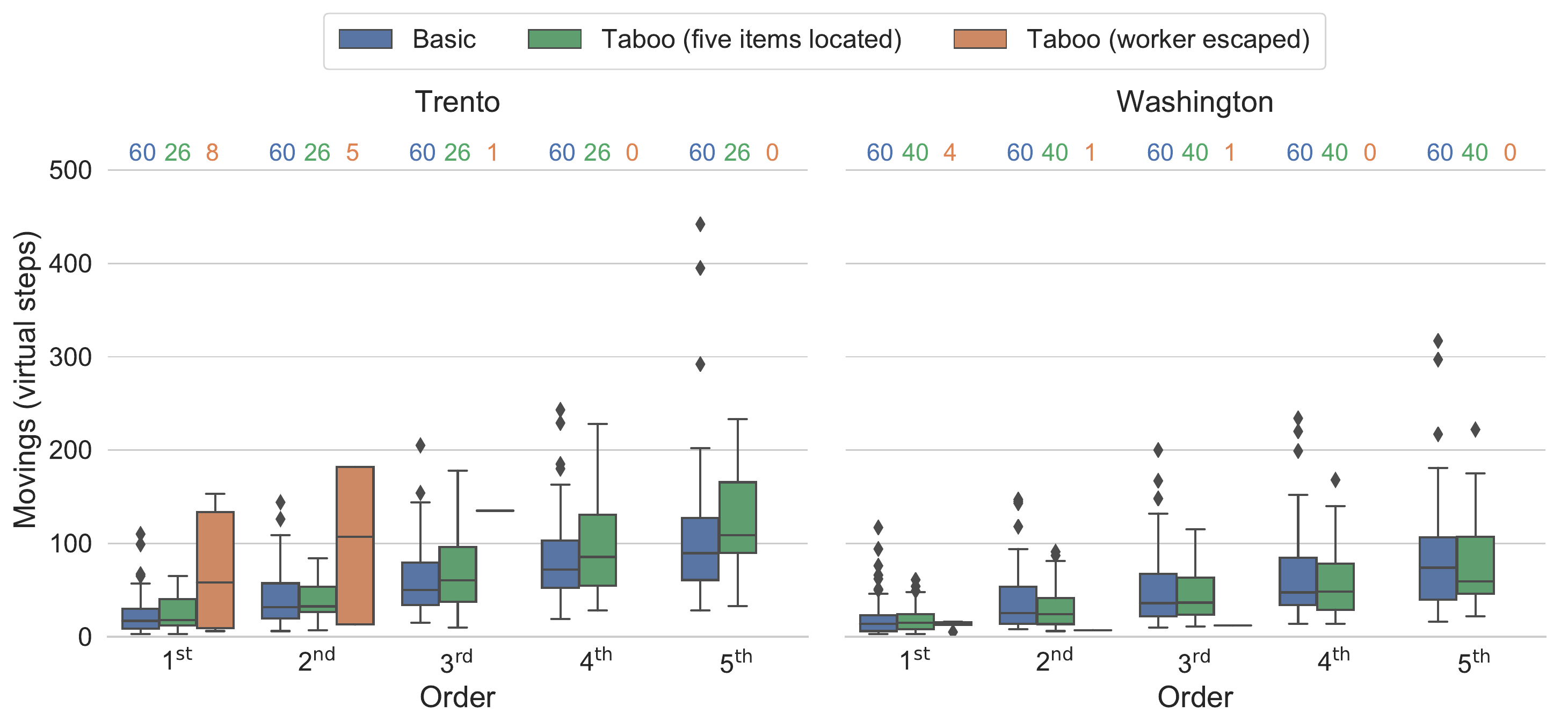}
 \caption{Number of steps per worker before submitting a bike rack during a task execution for each strategy: Basic (blue, leftmost boxplots), Taboo for workers that completed the task (green, center boxplots), and Taboo for workers that escaped (orange, rightmost boxplots)}
\label{figure:partial_movings}
\end{figure}

Figure \ref{figure:partial_movings} shows the analysis of the number of virtual steps each worker took after detecting a bike rack\footnote{Recall from section \ref{subsec:taskengine} that we delete the action log of workers that abandon the task in Basic}. The numbers on top of each boxplot represent the number of workers that submitted a 1st, 2nd, 3rd, 4th and 5th bike rack. For the Basic and Taboo for workers that completed the task, the number is always the same, but for Taboo for workers that escaped, this gives the number of workers that escaped before submitting that number of bike racks. For the case of Trento, we note an increase in the number of steps for Taboo with respect to Basic, which is very sharp for workers that triggered the escape condition. Also, the great majority of workers that escaped did so before finding a 3rd bike rack, 26 (out of 34) did so before submitting their first bike rack. No escaping workers ever found a 4th bike rack, with 20 workers escaping before submitting any bike rack.  For the case of Washington, we observe that almost all workers (16 out of 20) escaped before submitting their first bike rack. 

\begin{figure}
\includegraphics[width=0.75\linewidth,height=0.5\textheight]{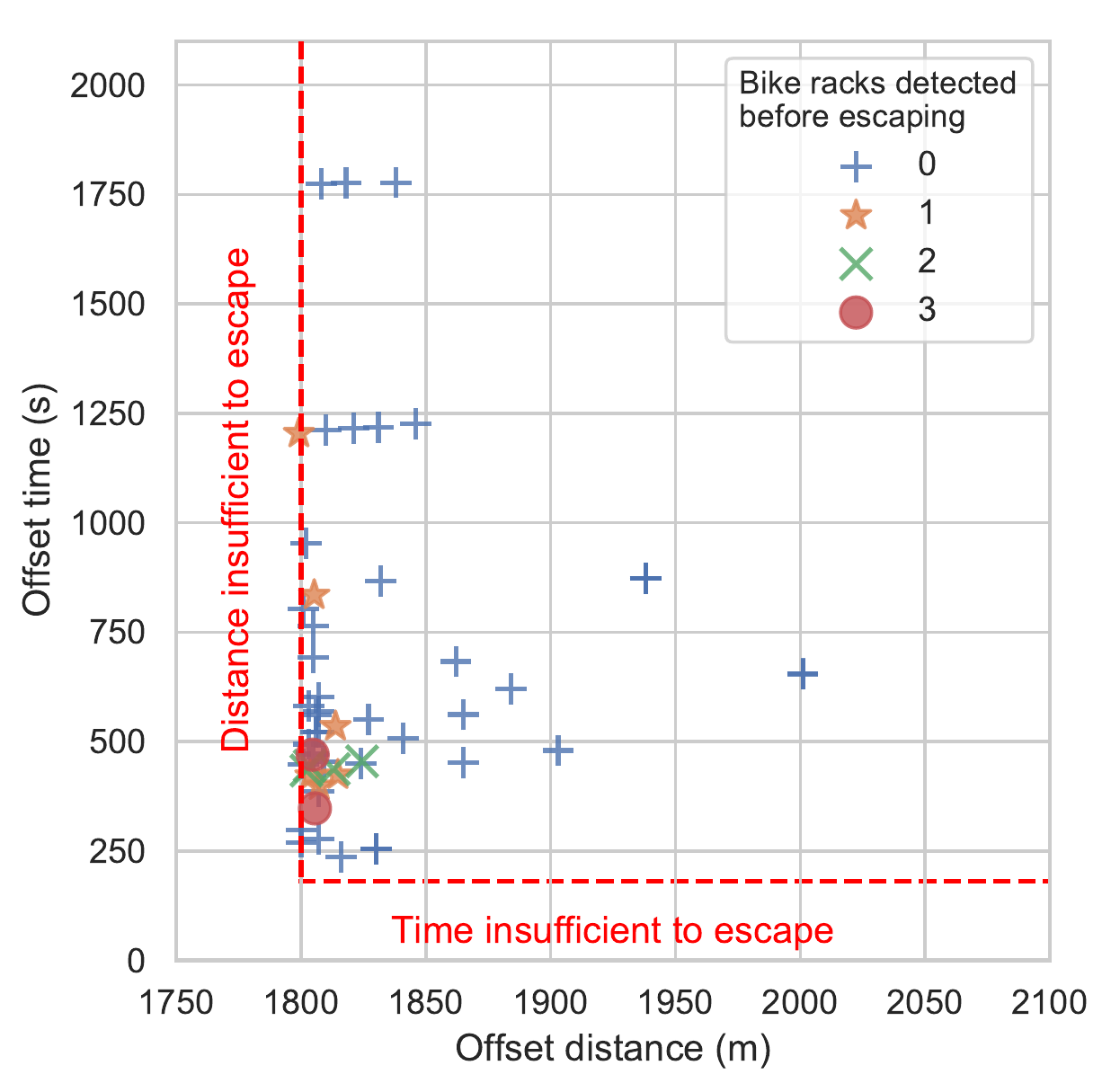}
\caption{Comparison of escape sub-conditions. Most escaping workers first triggered the 3 minutes time condition before the 1800m distance condition}
\label{figure:escape_strategy}
\end{figure}

Finally, to provide insight about our choice of escape time and distance, we extracted from the action log of workers that escaped in both areas the distance they walked, the time elapsed since their last detection, and how many bike racks were detected before escaping. Figure \ref{figure:escape_strategy} shows the results. the vertical dotted line signals the escape distance, while the horizontal dotted line signals the escape time. The concentration of points along the escape distance line indicates that most workers spent significantly more than 3 minutes without detecting any new bike rack before reaching the escape distance.
Intuitively, we expected all points being very close to either the escape time line or the escape distance line, however, there are some outliers that are not close to any of the lines. This is explained by the fact that we check the escape conditions only after the worker makes an action in the interface. It is then possible that a worker who has already reached the time condition, for a while, triggers the distance condition by taking a 200 meters step, ending up away from both lines in our chart.

\section{Worker interaction and behaviour}
\label{sec:behaviour}

In this section, we aim to answer RQ5: How much time do they spend on the task and how much distance they cover? Does the taboo strategy affect time spent and distance walked? Which are the most frequent errors that affect the user experience?

\begin{figure}

\includegraphics[width=0.8\linewidth]{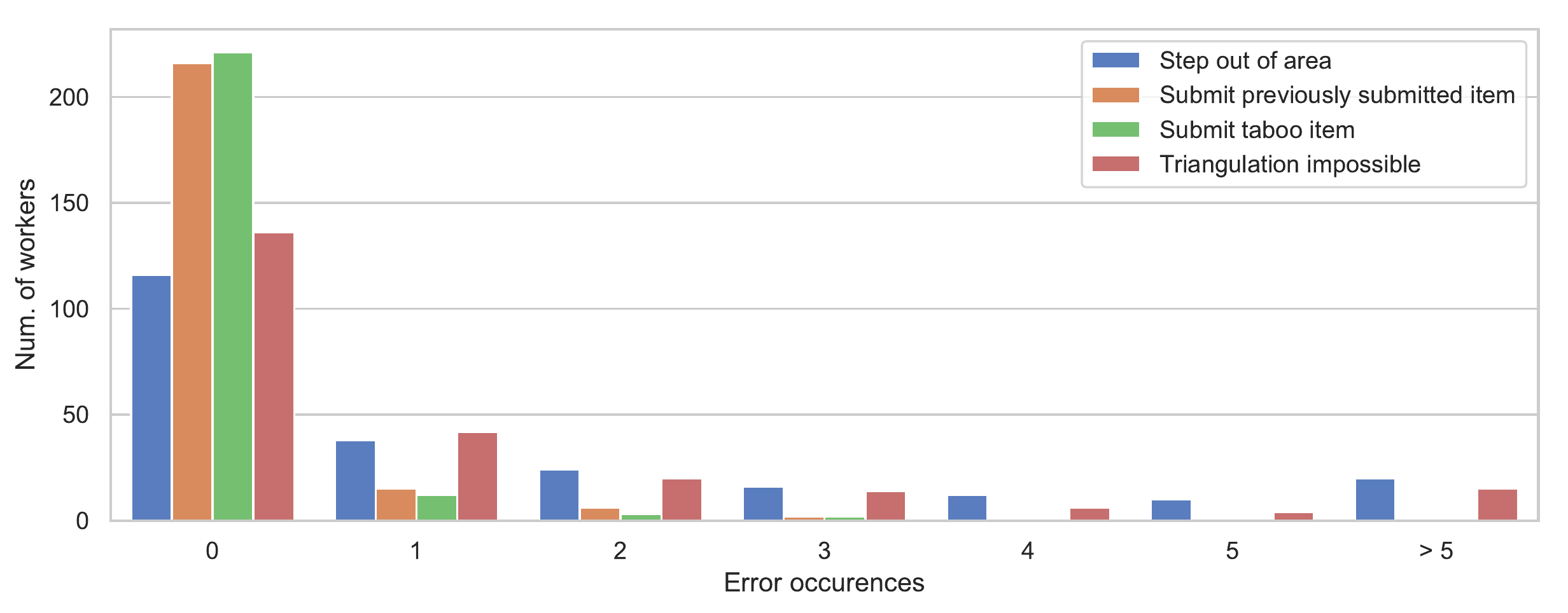}
\caption{Number of workers that triggered 0 to >5 interface errors}
\label{fig:errors}
\end{figure}

We counted from the action log the number of times workers triggered an interface error, namely, step out of the area boundary, submit the same PoI more than once, submit a PoI labeled as \emph{taboo}, and invalid submissions as for the triangulation method (cf. Section~\ref{sec:systemdesign}). Figure \ref{fig:errors} reveals that most workers were able to complete the task without triggering any error, or only once. The most common errors are stepping out of the area and failing the triangulation check, suggesting two areas of improvement to the interface. We also note that very few workers attempted to submit previously detected or taboo PoIs.

\begin{figure}
\centering
\includegraphics[width=\linewidth]{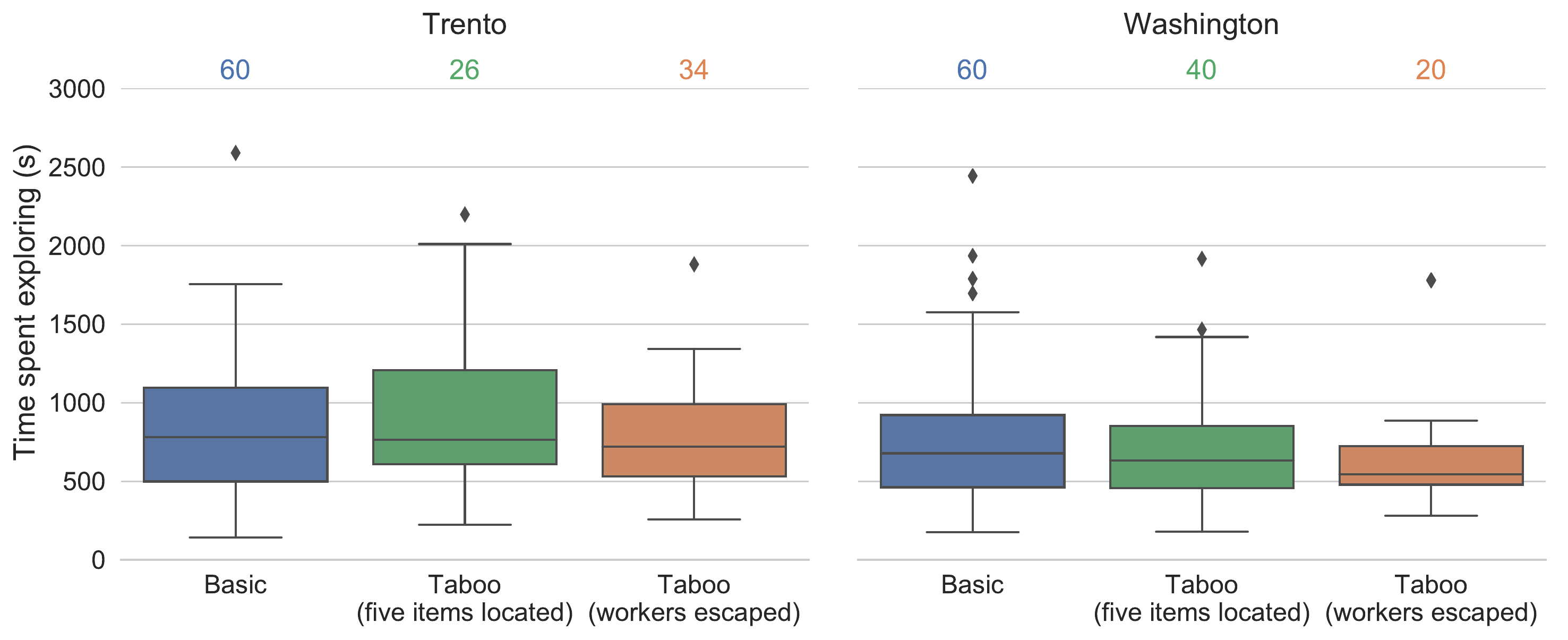}
 \caption{Distribution of time spent by workers for each strategy. Data for Taboo strategy is partitioned in workers that completed the task and workers that escaped the task.}
\label{figure:times}
\end{figure}

With respect to time to complete the task, we partitioned the measurements in three subsets per area of interest, as shown in figure~\ref{figure:times}. Time spent by workers that worked under Basic, that worked under Taboo and completed the task, and workers under Taboo that triggered the escape condition. All workers spent at least 7 minutes on the task, with one outlier spending more than 40 in Trento. All partitions in Trento have approximately the same median of around 750 seconds. For Washington, Basic and Taboo completed have similar medians, but for escaping workers, there is a difference of nearly 3 minutes less. These results suggest that the Taboo strategy does not significantly impact the amount of time workers spend in completing the task. 

\begin{figure}
\centering
\includegraphics[width=\linewidth]{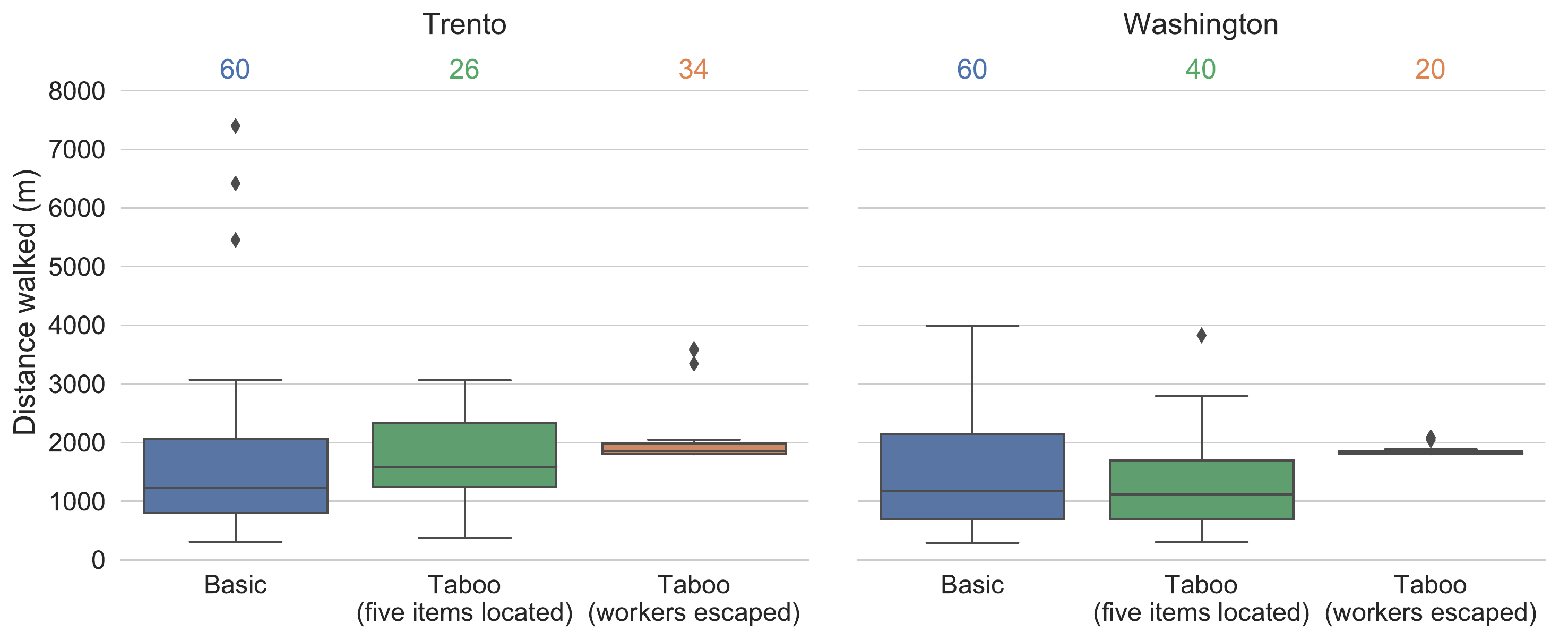}
 \caption{Distribution of distance covered by workers for each strategy. Data for Taboo strategy is partitioned in workers that completed the task and workers that escaped the task}
\label{figure:distances}
\end{figure}

With respect to walked distance, we partitioned data in the same way as we did with time (Figure~\ref{figure:distances}). In the case of Trento, there is a noticeable increase of approximately 30\% in the median distance walked under the Taboo strategy with respect to Basic, suggesting that the former does represents an impact. Interestingly, in combination with insight from Figure \ref{figure:times}, this means that despite covering a larger distance, workers under Taboo did not spend much more time than those under Basic, suggesting that either they explored faster after some time in the task, or that they retraced their path with longer steps. In Washington's case, there is no significant difference in the median walked distance between Basic and Taboo (completed). We believe this is due to the high density of bike racks in Washington, allowing workers to find close alternatives to taboo items. Workers that ended up escaping did walk significantly more than the others.

%
%

\section{Conclusions and Future Work}
\label{sec:conclusions}

In this paper, we have described the design, implementation, and evaluation of the Virtual City Explorer, a system to collect geo-spatial information through the generation of an exploration interface on top of street view imagery services, through which paid crowdworkers can report location of PoIs. The evaluation of our system focused on the task of generating maps of PoIs within a bounded area. We compared the maps generated with our approach against maps generated by an expert on the field, by volunteers in Open Street Maps, and by volunteers using an specific mobile app (Rackspotter) sponsored by a public authority. We also studied the effectiveness of making  \emph{Taboo} PoIs already detected by a number of workers, to encourage further exploration to improve coverage and completeness. Our Research Questions drove the comparison along several dimensions:

\textbf{Feasibility and completeness (RQ1 and RQ4):} Our results suggest that our approach has a low rate of false positives (things not an instance of the sought type of PoI). Crowdworkers do make mistakes individually, but to the aggregation component prevents them to appear in the final result, improving accuracy. 
In terms of completeness, we found that no approach dominates others. The VCE detected less PoIs overall, but it was able to detect some that were not detected by on-the-field approaches, suggesting that they might be complementary. When using the Taboo optimisation we observed the number of detected PoIs increase $30\%$ in the Trento area, but less than $10\%$ in the Washington area. In the latter case, there is a significant dissimilarity between the maps collected with the Basic and the Taboo strategy. All this suggests that the number of task executions and the number of per worker instances to find need to be adjusted according to the expected number of PoIs, or even, dynamically according to the coverage obtained so far.

\textbf{Area coverage (RQ2 and RQ4):}
In our experiment setting with 60 task executions, our approach achieved almost $90\%$ coverage of the Trento area, and almost $100\%$ of the Washington area. In the case of Trento, most of the unexplored bits corresponded to segments of street close to the boundary. When using the Taboo strategy, workers explored more, but roughly the same segments of street remained unexplored, suggesting that workers might need to be guided to them, for example, by making incoming workers start exploration in unexplored segments.

\textbf{Cost and generation time (RQ3 and RQ4):} Our approach is more expensive than free volunteered alternatives, but more than 5 times less expensive than an on-field survey. In turn, the on-field survey was much faster than our approach, for the studied area, but for larger areas where sending experts is too expensive, we believe our approach will be faster. For OSM maps, the time between first and last contribution was in the order of years, with less than 5 contributors, a consequence of the lack of control of number of contributors. We also analysed the association between number of task executions and detected PoIs, finding resemblance to a logarithmic curve, therefore, suggesting that for scenarios where the goal is to map a fixed number of PoIs (instead of of all of them), the number of executions could be reduced. We also found that the Taboo optimisation generates more confirmed PoIs with less detections, at the same cost than without it.

\textbf{Behaviour and learning effect (RQ5):} We did not find any significant variation in the amount of time spent in the task by workers in the Taboo and Basic strategies, suggesting that the Taboo strategy did not increase the complexity of the task for workers. This also suggests that workers starting exploring faster after finding some taboo bike racks. We hypothesize that this is due to them retracing their steps at a faster pace. In terms of struggle with the interface, we found that a significant majority of workers completed (or escaped) the task without committing any error, suggesting that the learning curve of our task is rather gentle.

In summary, the use of crowdworkers in a virtual space is a cost-effective alternative to paid on-site PoI mapping in regions where there is an absence of a dedicated volunteer mapping community, and can provide better direct control of coverage and a quicker turnaround time of results.

As future work, we would like to compare our free exploration approach with alternatives that guide the worker to explore a fixed segment of street (similar to~\cite{saha_pilot_2017}), and assigning starting points for incoming workers in less explored areas. We would also like to explore dynamical reward strategies, for example, based on the number of PoIs reported. Finally, in real-world applications, there is often no other reference datasets available, therefore, we would like to implement measures to assess agreement among workers to guarantee a satisfying quality level of reported PoIs.

\textbf{Acknowledgements:}
We would like to thank the crowd workers who participated in this work, and the team at Municipality of Trento for kindly answering questions about their bike rack map. The Rackspotter team also answered several enquiries about their dataset. This study was partially supported by the \emph{QROWD} project, part of the European Union’s Horizon 2020 research and innovation programme, under grant agreement No 723088.


\bibliographystyle{ACM-Reference-Format}
\bibliography{TIST-VCE}


\begin{thebibliography}{23}


\ifx \showCODEN    \undefined \def \showCODEN     #1{\unskip}     \fi
\ifx \showDOI      \undefined \def \showDOI       #1{#1}\fi
\ifx \showISBNx    \undefined \def \showISBNx     #1{\unskip}     \fi
\ifx \showISBNxiii \undefined \def \showISBNxiii  #1{\unskip}     \fi
\ifx \showISSN     \undefined \def \showISSN      #1{\unskip}     \fi
\ifx \showLCCN     \undefined \def \showLCCN      #1{\unskip}     \fi
\ifx \shownote     \undefined \def \shownote      #1{#1}          \fi
\ifx \showarticletitle \undefined \def \showarticletitle #1{#1}   \fi
\ifx \showURL      \undefined \def \showURL       {\relax}        \fi
\providecommand\bibfield[2]{#2}
\providecommand\bibinfo[2]{#2}
\providecommand\natexlab[1]{#1}
\providecommand\showeprint[2][]{arXiv:#2}

\bibitem[\protect\citeauthoryear{Anguelov, Dulong, Filip, Frueh, Lafon, Lyon,
  Ogale, Vincent, and Weaver}{Anguelov et~al\mbox{.}}{2010}]%
        {anguelov_google_2010}
\bibfield{author}{\bibinfo{person}{Dragomir Anguelov}, \bibinfo{person}{Carole
  Dulong}, \bibinfo{person}{Daniel Filip}, \bibinfo{person}{Christian Frueh},
  \bibinfo{person}{Stéphane Lafon}, \bibinfo{person}{Richard Lyon},
  \bibinfo{person}{Abhijit Ogale}, \bibinfo{person}{Luc Vincent}, {and}
  \bibinfo{person}{Josh Weaver}.} \bibinfo{year}{2010}\natexlab{}.
\newblock \showarticletitle{Google {Street} {View}: {Capturing} the {World} at
  {Street} {Level}}.
\newblock \bibinfo{journal}{\emph{Computer}} \bibinfo{volume}{43},
  \bibinfo{number}{6} (\bibinfo{date}{June} \bibinfo{year}{2010}),
  \bibinfo{pages}{32--38}.
\newblock
\showISSN{0018-9162}
\urldef\tempurl%
\url{https://doi.org/10.1109/MC.2010.170}
\showDOI{\tempurl}


\bibitem[\protect\citeauthoryear{Badland, Opit, Witten, Kearns, and
  Mavoa}{Badland et~al\mbox{.}}{2010}]%
        {badland_can_2010}
\bibfield{author}{\bibinfo{person}{Hannah~M. Badland}, \bibinfo{person}{Simon
  Opit}, \bibinfo{person}{Karen Witten}, \bibinfo{person}{Robin~A. Kearns},
  {and} \bibinfo{person}{Suzanne Mavoa}.} \bibinfo{year}{2010}\natexlab{}.
\newblock \showarticletitle{Can {Virtual} {Streetscape} {Audits} {Reliably}
  {Replace} {Physical} {Streetscape} {Audits}?}
\newblock \bibinfo{journal}{\emph{Journal of Urban Health}}
  \bibinfo{volume}{87}, \bibinfo{number}{6} (\bibinfo{date}{Dec.}
  \bibinfo{year}{2010}), \bibinfo{pages}{1007--1016}.
\newblock
\showISSN{1099-3460, 1468-2869}
\urldef\tempurl%
\url{https://doi.org/10.1007/s11524-010-9505-x}
\showDOI{\tempurl}


\bibitem[\protect\citeauthoryear{Dorn, Törnros, and Zipf}{Dorn
  et~al\mbox{.}}{2015}]%
        {dorn_quality_2015}
\bibfield{author}{\bibinfo{person}{Helen Dorn}, \bibinfo{person}{Tobias
  Törnros}, {and} \bibinfo{person}{Alexander Zipf}.}
  \bibinfo{year}{2015}\natexlab{}.
\newblock \showarticletitle{Quality {Evaluation} of {VGI} {Using}
  {Authoritative} {Data}—{A} {Comparison} with {Land} {Use} {Data} in
  {Southern} {Germany}}.
\newblock \bibinfo{journal}{\emph{ISPRS International Journal of
  Geo-Information}} \bibinfo{volume}{4}, \bibinfo{number}{3}
  (\bibinfo{date}{Sept.} \bibinfo{year}{2015}), \bibinfo{pages}{1657--1671}.
\newblock
\urldef\tempurl%
\url{https://doi.org/10.3390/ijgi4031657}
\showDOI{\tempurl}


\bibitem[\protect\citeauthoryear{Fan, Zipf, Fu, and Neis}{Fan
  et~al\mbox{.}}{2014}]%
        {fan_quality_2014}
\bibfield{author}{\bibinfo{person}{Hongchao Fan}, \bibinfo{person}{Alexander
  Zipf}, \bibinfo{person}{Qing Fu}, {and} \bibinfo{person}{Pascal Neis}.}
  \bibinfo{year}{2014}\natexlab{}.
\newblock \showarticletitle{Quality assessment for building footprints data on
  {OpenStreetMap}}.
\newblock \bibinfo{journal}{\emph{International Journal of Geographical
  Information Science}} \bibinfo{volume}{28}, \bibinfo{number}{4}
  (\bibinfo{date}{April} \bibinfo{year}{2014}), \bibinfo{pages}{700--719}.
\newblock
\showISSN{1365-8816}
\urldef\tempurl%
\url{https://doi.org/10.1080/13658816.2013.867495}
\showDOI{\tempurl}


\bibitem[\protect\citeauthoryear{Goodchild, Fu, and Rich}{Goodchild
  et~al\mbox{.}}{2007}]%
        {goodchild_sharing_2007}
\bibfield{author}{\bibinfo{person}{Michael~F. Goodchild},
  \bibinfo{person}{Pinde Fu}, {and} \bibinfo{person}{Paul Rich}.}
  \bibinfo{year}{2007}\natexlab{}.
\newblock \showarticletitle{Sharing {Geographic} {Information}: {An}
  {Assessment} of the {Geospatial} {One}-{Stop}}.
\newblock \bibinfo{journal}{\emph{Annals of the Association of American
  Geographers}} \bibinfo{volume}{97}, \bibinfo{number}{2} (\bibinfo{date}{June}
  \bibinfo{year}{2007}), \bibinfo{pages}{250--266}.
\newblock
\showISSN{0004-5608, 1467-8306}
\urldef\tempurl%
\url{https://doi.org/10.1111/j.1467-8306.2007.00534.x}
\showDOI{\tempurl}


\bibitem[\protect\citeauthoryear{Griew, Hillsdon, Foster, Coombes, Jones, and
  Wilkinson}{Griew et~al\mbox{.}}{2013}]%
        {griew_developing_2013}
\bibfield{author}{\bibinfo{person}{Pippa Griew}, \bibinfo{person}{Melvyn
  Hillsdon}, \bibinfo{person}{Charlie Foster}, \bibinfo{person}{Emma Coombes},
  \bibinfo{person}{Andy Jones}, {and} \bibinfo{person}{Paul Wilkinson}.}
  \bibinfo{year}{2013}\natexlab{}.
\newblock \showarticletitle{Developing and testing a street audit tool using
  {Google} {Street} {View} to measure environmental supportiveness for physical
  activity}.
\newblock \bibinfo{journal}{\emph{International Journal of Behavioral Nutrition
  and Physical Activity}} \bibinfo{volume}{10}, \bibinfo{number}{1}
  (\bibinfo{year}{2013}), \bibinfo{pages}{103}.
\newblock
\showISSN{1479-5868}
\urldef\tempurl%
\url{https://doi.org/10.1186/1479-5868-10-103}
\showDOI{\tempurl}


\bibitem[\protect\citeauthoryear{Haklay and Weber}{Haklay and Weber}{2008}]%
        {haklay_openstreetmap:_2008}
\bibfield{author}{\bibinfo{person}{M. Haklay} {and} \bibinfo{person}{P.
  Weber}.} \bibinfo{year}{2008}\natexlab{}.
\newblock \showarticletitle{{OpenStreetMap}: {User}-{Generated} {Street}
  {Maps}}.
\newblock \bibinfo{journal}{\emph{IEEE Pervasive Computing}}
  \bibinfo{volume}{7}, \bibinfo{number}{4} (\bibinfo{date}{Oct.}
  \bibinfo{year}{2008}), \bibinfo{pages}{12--18}.
\newblock
\showISSN{1536-1268}
\urldef\tempurl%
\url{https://doi.org/10.1109/MPRV.2008.80}
\showDOI{\tempurl}


\bibitem[\protect\citeauthoryear{Haklay, Basiouka, Antoniou, and Ather}{Haklay
  et~al\mbox{.}}{2010}]%
        {haklay_how_2010}
\bibfield{author}{\bibinfo{person}{Mordechai~(Muki) Haklay},
  \bibinfo{person}{Sofia Basiouka}, \bibinfo{person}{Vyron Antoniou}, {and}
  \bibinfo{person}{Aamer Ather}.} \bibinfo{year}{2010}\natexlab{}.
\newblock \showarticletitle{How {Many} {Volunteers} {Does} it {Take} to {Map}
  an {Area} {Well}? {The} {Validity} of {Linus}’ {Law} to {Volunteered}
  {Geographic} {Information}}.
\newblock \bibinfo{journal}{\emph{The Cartographic Journal}}
  \bibinfo{volume}{47}, \bibinfo{number}{4} (\bibinfo{date}{Nov.}
  \bibinfo{year}{2010}), \bibinfo{pages}{315--322}.
\newblock
\showISSN{0008-7041, 1743-2774}
\urldef\tempurl%
\url{https://doi.org/10.1179/000870410X12911304958827}
\showDOI{\tempurl}


\bibitem[\protect\citeauthoryear{Hara, Froehlich, Azenkot, Campbell, Bennett,
  Le, Pannella, Moore, Minckler, and Ng}{Hara et~al\mbox{.}}{2015}]%
        {hara_improving_2015}
\bibfield{author}{\bibinfo{person}{Kotaro Hara}, \bibinfo{person}{Jon~E.
  Froehlich}, \bibinfo{person}{Shiri Azenkot}, \bibinfo{person}{Megan
  Campbell}, \bibinfo{person}{Cynthia~L. Bennett}, \bibinfo{person}{Vicki Le},
  \bibinfo{person}{Sean Pannella}, \bibinfo{person}{Robert Moore},
  \bibinfo{person}{Kelly Minckler}, {and} \bibinfo{person}{Rochelle~H. Ng}.}
  \bibinfo{year}{2015}\natexlab{}.
\newblock \showarticletitle{Improving {Public} {Transit} {Accessibility} for
  {Blind} {Riders} by {Crowdsourcing} {Bus} {Stop} {Landmark} {Locations} with
  {Google} {Street} {View}: {An} {Extended} {Analysis}}.
\newblock \bibinfo{journal}{\emph{ACM Transactions on Accessible Computing}}
  \bibinfo{volume}{6}, \bibinfo{number}{2} (\bibinfo{date}{March}
  \bibinfo{year}{2015}), \bibinfo{pages}{1--23}.
\newblock
\showISSN{19367228}
\urldef\tempurl%
\url{https://doi.org/10.1145/2717513}
\showDOI{\tempurl}


\bibitem[\protect\citeauthoryear{Hara, Le, and Froehlich}{Hara
  et~al\mbox{.}}{2013}]%
        {hara_combining_2013}
\bibfield{author}{\bibinfo{person}{Kotaro Hara}, \bibinfo{person}{Vicki Le},
  {and} \bibinfo{person}{Jon Froehlich}.} \bibinfo{year}{2013}\natexlab{}.
\newblock \showarticletitle{Combining crowdsourcing and google street view to
  identify street-level accessibility problems}. \bibinfo{publisher}{ACM
  Press}, \bibinfo{pages}{631--640}.
\newblock
\showISBNx{978-1-4503-1899-0}
\urldef\tempurl%
\url{https://doi.org/10.1145/2470654.2470744}
\showDOI{\tempurl}


\bibitem[\protect\citeauthoryear{Hardy, Frew, and Goodchild}{Hardy
  et~al\mbox{.}}{2012}]%
        {hardy_volunteered_2012}
\bibfield{author}{\bibinfo{person}{Darren Hardy}, \bibinfo{person}{James Frew},
  {and} \bibinfo{person}{Michael~F. Goodchild}.}
  \bibinfo{year}{2012}\natexlab{}.
\newblock \showarticletitle{Volunteered geographic information production as a
  spatial process}.
\newblock \bibinfo{journal}{\emph{International Journal of Geographical
  Information Science}} \bibinfo{volume}{26}, \bibinfo{number}{7}
  (\bibinfo{date}{July} \bibinfo{year}{2012}), \bibinfo{pages}{1191--1212}.
\newblock
\showISSN{1365-8816, 1362-3087}
\urldef\tempurl%
\url{https://doi.org/10.1080/13658816.2011.629618}
\showDOI{\tempurl}


\bibitem[\protect\citeauthoryear{Hecht and Gergle}{Hecht and Gergle}{2010}]%
        {hecht_localness_2010}
\bibfield{author}{\bibinfo{person}{Brent~J. Hecht} {and}
  \bibinfo{person}{Darren Gergle}.} \bibinfo{year}{2010}\natexlab{}.
\newblock \showarticletitle{On the "{Localness}" of {User}-generated
  {Content}}. In \bibinfo{booktitle}{\emph{Proceedings of the 2010 {ACM}
  {Conference} on {Computer} {Supported} {Cooperative} {Work}}}
  \emph{(\bibinfo{series}{{CSCW} '10})}. \bibinfo{publisher}{ACM},
  \bibinfo{address}{New York, NY, USA}, \bibinfo{pages}{229--232}.
\newblock
\showISBNx{978-1-60558-795-0}
\urldef\tempurl%
\url{https://doi.org/10.1145/1718918.1718962}
\showDOI{\tempurl}


\bibitem[\protect\citeauthoryear{Jiang, Kresin, Bregt, Kooistra, Pareschi, van
  Putten, Volten, and Wesseling}{Jiang et~al\mbox{.}}{2016}]%
        {jiang_citizen_2016}
\bibfield{author}{\bibinfo{person}{Qijun Jiang}, \bibinfo{person}{Frank
  Kresin}, \bibinfo{person}{Arnold~K. Bregt}, \bibinfo{person}{Lammert
  Kooistra}, \bibinfo{person}{Emma Pareschi}, \bibinfo{person}{Edith van
  Putten}, \bibinfo{person}{Hester Volten}, {and} \bibinfo{person}{Joost
  Wesseling}.} \bibinfo{year}{2016}\natexlab{}.
\newblock \bibinfo{title}{Citizen {Sensing} for {Improved} {Urban}
  {Environmental} {Monitoring}}.
\newblock
\newblock
\urldef\tempurl%
\url{https://doi.org/10.1155/2016/5656245}
\showDOI{\tempurl}


\bibitem[\protect\citeauthoryear{Kandappu, Misra, Cheng, Tandriansyah, and
  Lau}{Kandappu et~al\mbox{.}}{2018}]%
        {kandappu_obfuscation_2018}
\bibfield{author}{\bibinfo{person}{Thivya Kandappu}, \bibinfo{person}{Archan
  Misra}, \bibinfo{person}{Shih-Fen Cheng}, \bibinfo{person}{Randy
  Tandriansyah}, {and} \bibinfo{person}{Hoong~Chuin Lau}.}
  \bibinfo{year}{2018}\natexlab{}.
\newblock \showarticletitle{Obfuscation {At}-{Source}: {Privacy} in
  {Context}-{Aware} {Mobile} {Crowd}-{Sourcing}}.
\newblock \bibinfo{journal}{\emph{Proc. ACM Interact. Mob. Wearable Ubiquitous
  Technol.}} \bibinfo{volume}{2}, \bibinfo{number}{1} (\bibinfo{date}{March}
  \bibinfo{year}{2018}), \bibinfo{pages}{16:1--16:24}.
\newblock
\showISSN{2474-9567}
\urldef\tempurl%
\url{https://doi.org/10.1145/3191748}
\showDOI{\tempurl}


\bibitem[\protect\citeauthoryear{Mooney and Corcoran}{Mooney and
  Corcoran}{2014}]%
        {mooney_analysis_2014}
\bibfield{author}{\bibinfo{person}{Peter Mooney} {and} \bibinfo{person}{Padraig
  Corcoran}.} \bibinfo{year}{2014}\natexlab{}.
\newblock \showarticletitle{Analysis of {Interaction} and {Co}-editing
  {Patterns} amongst {OpenStreetMap} {Contributors}}.
\newblock \bibinfo{journal}{\emph{Transactions in GIS}} \bibinfo{volume}{18},
  \bibinfo{number}{5} (\bibinfo{date}{Oct.} \bibinfo{year}{2014}),
  \bibinfo{pages}{633--659}.
\newblock
\showISSN{1467-9671}
\urldef\tempurl%
\url{https://doi.org/10.1111/tgis.12051}
\showDOI{\tempurl}


\bibitem[\protect\citeauthoryear{Quattrone, Dittus, and Capra}{Quattrone
  et~al\mbox{.}}{2017}]%
        {quattrone_work_2017}
\bibfield{author}{\bibinfo{person}{Giovanni Quattrone}, \bibinfo{person}{Martin
  Dittus}, {and} \bibinfo{person}{Licia Capra}.}
  \bibinfo{year}{2017}\natexlab{}.
\newblock \showarticletitle{Work {Always} in {Progress}: {Analysing}
  {Maintenance} {Practices} in {Spatial} {Crowd}-sourced {Datasets}}. In
  \bibinfo{booktitle}{\emph{Proceedings of the 2017 {ACM} {Conference} on
  {Computer} {Supported} {Cooperative} {Work} and {Social} {Computing}}}
  \emph{(\bibinfo{series}{{CSCW} '17})}. \bibinfo{publisher}{ACM},
  \bibinfo{address}{New York, NY, USA}, \bibinfo{pages}{1876--1889}.
\newblock
\showISBNx{978-1-4503-4335-0}
\urldef\tempurl%
\url{https://doi.org/10.1145/2998181.2998267}
\showDOI{\tempurl}


\bibitem[\protect\citeauthoryear{Ruiz-Correa, Santani, Ramírez-Salazar,
  Ruiz-Correa, Rendón-Huerta, Olmos-Carrillo, Sandoval-Mexicano,
  Arcos-García, Hasimoto-Beltrán, and Gatica-Perez}{Ruiz-Correa
  et~al\mbox{.}}{2017}]%
        {ruiz-correa_sensecityvity:_2017}
\bibfield{author}{\bibinfo{person}{S. Ruiz-Correa}, \bibinfo{person}{D.
  Santani}, \bibinfo{person}{B. Ramírez-Salazar}, \bibinfo{person}{I.
  Ruiz-Correa}, \bibinfo{person}{F.~A. Rendón-Huerta}, \bibinfo{person}{C.
  Olmos-Carrillo}, \bibinfo{person}{B.~C. Sandoval-Mexicano},
  \bibinfo{person}{Á~H. Arcos-García}, \bibinfo{person}{R.
  Hasimoto-Beltrán}, {and} \bibinfo{person}{D. Gatica-Perez}.}
  \bibinfo{year}{2017}\natexlab{}.
\newblock \showarticletitle{{SenseCityVity}: {Mobile} {Crowdsourcing}, {Urban}
  {Awareness}, and {Collective} {Action} in {Mexico}}.
\newblock \bibinfo{journal}{\emph{IEEE Pervasive Computing}}
  \bibinfo{volume}{16}, \bibinfo{number}{2} (\bibinfo{date}{April}
  \bibinfo{year}{2017}), \bibinfo{pages}{44--53}.
\newblock
\showISSN{1536-1268}
\urldef\tempurl%
\url{https://doi.org/10.1109/MPRV.2017.32}
\showDOI{\tempurl}


\bibitem[\protect\citeauthoryear{Saha, Hara, Behnezhad, Li, Saugstad, Maddali,
  Chen, and Froehlich}{Saha et~al\mbox{.}}{2017}]%
        {saha_pilot_2017}
\bibfield{author}{\bibinfo{person}{Manaswi Saha}, \bibinfo{person}{Kotaro
  Hara}, \bibinfo{person}{Soheil Behnezhad}, \bibinfo{person}{Anthony Li},
  \bibinfo{person}{Michael Saugstad}, \bibinfo{person}{Hanuma Maddali},
  \bibinfo{person}{Sage Chen}, {and} \bibinfo{person}{Jon~E. Froehlich}.}
  \bibinfo{year}{2017}\natexlab{}.
\newblock \showarticletitle{A {Pilot} {Deployment} of an {Online} {Tool} for
  {Large}-{Scale} {Virtual} {Auditing} of {Urban} {Accessibility}}. In
  \bibinfo{booktitle}{\emph{Proceedings of the 19th {International} {ACM}
  {SIGACCESS} {Conference} on {Computers} and {Accessibility}}}.
  \bibinfo{publisher}{ACM Press}, \bibinfo{pages}{305--306}.
\newblock
\showISBNx{978-1-4503-4926-0}
\urldef\tempurl%
\url{https://doi.org/10.1145/3132525.3134775}
\showDOI{\tempurl}


\bibitem[\protect\citeauthoryear{Schubert, Sander, Ester, Kriegel, and
  Xu}{Schubert et~al\mbox{.}}{2017}]%
        {schubert_dbscan_2017}
\bibfield{author}{\bibinfo{person}{Erich Schubert}, \bibinfo{person}{Jörg
  Sander}, \bibinfo{person}{Martin Ester}, \bibinfo{person}{Hans~Peter
  Kriegel}, {and} \bibinfo{person}{Xiaowei Xu}.}
  \bibinfo{year}{2017}\natexlab{}.
\newblock \showarticletitle{{DBSCAN} {Revisited}, {Revisited}: {Why} and {How}
  {You} {Should} ({Still}) {Use} {DBSCAN}}.
\newblock \bibinfo{journal}{\emph{ACM Trans. Database Syst.}}
  \bibinfo{volume}{42}, \bibinfo{number}{3} (\bibinfo{date}{July}
  \bibinfo{year}{2017}), \bibinfo{pages}{19:1--19:21}.
\newblock
\showISSN{0362-5915}
\urldef\tempurl%
\url{https://doi.org/10.1145/3068335}
\showDOI{\tempurl}


\bibitem[\protect\citeauthoryear{Sui, Elwood, and Goodchild}{Sui
  et~al\mbox{.}}{2012}]%
        {sui_crowdsourcing_2012}
\bibfield{author}{\bibinfo{person}{Daniel Sui}, \bibinfo{person}{Sarah Elwood},
  {and} \bibinfo{person}{Michael Goodchild}.} \bibinfo{year}{2012}\natexlab{}.
\newblock \bibinfo{booktitle}{\emph{Crowdsourcing {Geographic} {Knowledge}:
  {Volunteered} {Geographic} {Information} ({VGI}) in {Theory} and
  {Practice}}}.
\newblock \bibinfo{publisher}{Springer Publishing Company, Incorporated}.
\newblock
\showISBNx{978-94-007-4586-5}


\bibitem[\protect\citeauthoryear{Tong, Wang, Zhou, Ding, Chen, Ye, and Xu}{Tong
  et~al\mbox{.}}{2017}]%
        {tong_flexible_2017}
\bibfield{author}{\bibinfo{person}{Yongxin Tong}, \bibinfo{person}{Libin Wang},
  \bibinfo{person}{Zimu Zhou}, \bibinfo{person}{Bolin Ding},
  \bibinfo{person}{Lei Chen}, \bibinfo{person}{Jieping Ye}, {and}
  \bibinfo{person}{Ke Xu}.} \bibinfo{year}{2017}\natexlab{}.
\newblock \showarticletitle{Flexible {Online} {Task} {Assignment} in
  {Real}-time {Spatial} {Data}}.
\newblock \bibinfo{journal}{\emph{Proc. VLDB Endow.}} \bibinfo{volume}{10},
  \bibinfo{number}{11} (\bibinfo{date}{Aug.} \bibinfo{year}{2017}),
  \bibinfo{pages}{1334--1345}.
\newblock
\showISSN{2150-8097}
\urldef\tempurl%
\url{https://doi.org/10.14778/3137628.3137643}
\showDOI{\tempurl}


\bibitem[\protect\citeauthoryear{Vanwolleghem, Van~Dyck, Ducheyne,
  De~Bourdeaudhuij, and Cardon}{Vanwolleghem et~al\mbox{.}}{2014}]%
        {vanwolleghem_assessing_2014}
\bibfield{author}{\bibinfo{person}{Griet Vanwolleghem},
  \bibinfo{person}{Delfien Van~Dyck}, \bibinfo{person}{Fabian Ducheyne},
  \bibinfo{person}{Ilse De~Bourdeaudhuij}, {and} \bibinfo{person}{Greet
  Cardon}.} \bibinfo{year}{2014}\natexlab{}.
\newblock \showarticletitle{Assessing the environmental characteristics of
  cycling routes to school: a study on the reliability and validity of a
  {Google} {Street} {View}-based audit}.
\newblock \bibinfo{journal}{\emph{International Journal of Health Geographics}}
  \bibinfo{volume}{13}, \bibinfo{number}{1} (\bibinfo{year}{2014}),
  \bibinfo{pages}{19}.
\newblock
\showISSN{1476-072X}
\urldef\tempurl%
\url{https://doi.org/10.1186/1476-072X-13-19}
\showDOI{\tempurl}


\bibitem[\protect\citeauthoryear{Zhao and Han}{Zhao and Han}{2016}]%
        {zhao_spatial_2016}
\bibfield{author}{\bibinfo{person}{Y. Zhao} {and} \bibinfo{person}{Q. Han}.}
  \bibinfo{year}{2016}\natexlab{}.
\newblock \showarticletitle{Spatial crowdsourcing: current state and future
  directions}.
\newblock \bibinfo{journal}{\emph{IEEE Communications Magazine}}
  \bibinfo{volume}{54}, \bibinfo{number}{7} (\bibinfo{date}{July}
  \bibinfo{year}{2016}), \bibinfo{pages}{102--107}.
\newblock
\showISSN{0163-6804}
\urldef\tempurl%
\url{https://doi.org/10.1109/MCOM.2016.7509386}
\showDOI{\tempurl}


\end{thebibliography}

\end{document}